\documentclass[11pt]{article}
\usepackage{jheppub}

\pdfoutput=1
\usepackage{amsmath,amssymb}
\usepackage{epsfig}
\usepackage{hyperref}
\usepackage{color}
\usepackage{slashed}
\usepackage{placeins}

\usepackage{amsfonts}
\usepackage{graphicx, rotating}
\usepackage{epstopdf}
\usepackage{latexsym}
\usepackage{rotating}

\usepackage[normalem]{ulem}

\usepackage{subfig}

\usepackage[export]{adjustbox}

\usepackage[shortlabels]{enumitem}

\usepackage[font={small}]{caption}   

\definecolor{myred}{rgb}{0.7, 0, 0}
\definecolor{myblue}{rgb}{0, 0, 0.7}
\definecolor{mygreen}{rgb}{0.04, 0.7, 0.5}

\hypersetup{colorlinks,citecolor=myred,linkcolor=myblue,urlcolor=myblue,linktocpage=true}

 \def\be   {\begin{equation}}   \def\ee   {\end{equation}}
 \def\ba   {\begin{array}}      \def\ea   {\end{array}}
 \def\bea  {\begin{eqnarray}}   \def\eea  {\end{eqnarray}}
 \def\bean {\begin{eqnarray*}}  \def\eean {\end{eqnarray*}}
 
 \def\bry{\begin{array}}
 \def\ery{\end{array}}

\numberwithin{equation}{section}

\begin{document}

\color{black}

\title{
Baryon Asymmetry from Electroweak-Symmetric Domain Walls
}

\author{Jacopo Azzola,}

\author{Oleksii Matsedonskyi,}

\author{Andreas Weiler}

\affiliation {Technische Universit\"at M\"unchen, Physik-Department, James-Franck-Strasse 1, 85748 Garching, Germany}

\emailAdd{jacopo.azzola@tum.de, oleksii.matsedonskyi@tum.de, andreas.weiler@tum.de}

\abstract{
We investigate electroweak baryogenesis from domain walls with electroweak-symmetric cores moving through the electroweak-broken plasma. In the thick-wall regime, CP-violating semiclassical forces generate chiral asymmetries that source baryon number through transport and weak sphaleron processes. We show that the baryon yield is governed by the hierarchy between the wall width, the CP-violating source width, and the diffusion length, and we identify the corresponding scaling behavior in the relevant parametric limits. A distinctive feature of this mechanism is the interference between the two faces of the domain wall, which leads to qualitatively different behavior for CP-violating sources that are even or odd under wall-orientation reversal. We construct a simplified description that captures these effects and reproduces the predictions of the full transport system in a broad range of parameter space. Applying our framework to a singlet-extended Standard Model, we delineate the region in which electroweak-symmetric domain walls can generate the observed baryon asymmetry.
}

\maketitle


\section{Introduction}

Electroweak baryogenesis (EWBG)~\cite{Shaposhnikov:1987tw,Cohen:1990it} (see e.g.~\cite{Morrissey:2012db,Cline:2006ts} for reviews) provides an attractive explanation for the matter--antimatter asymmetry of the Universe, with a strong potential for experimental tests through terrestrial experiments and cosmological observations. 
To satisfy the Sakharov conditions~\cite{Sakharov:1967dj}, EWBG exploits baryon-number violating processes that are present already in the SM, namely weak sphaleron transitions. 
The required departure from equilibrium is typically realized by the rapid motion of the interface between the broken and unbroken EW phases during a strongly first-order electroweak phase transition.

A comparatively less explored mechanism capable of generating similar out-of-equilibrium dynamics involves domain walls (DW) \cite{Brandenberger:1994mq,Abel:1995uc,Brandenberger:2005bx,Bai:2021xyf,Sassi:2024cyb,Schroder:2024gsi, Klipfel:2026nzx}. Such defects arise when the theory admits a disconnected set of vacua. A minimal realization is provided by a singlet-extended Standard Model~\cite{Azzola:2024pzq} containing a real scalar field $S$ invariant under the discrete transformation $S \to -S$. Spontaneous breaking of this symmetry leads to the formation of domains corresponding to the degenerate vacua $\langle S \rangle = \pm v_S$, separated by domain walls.

\begin{figure}[t]
\center
\includegraphics[width=0.8 \textwidth]{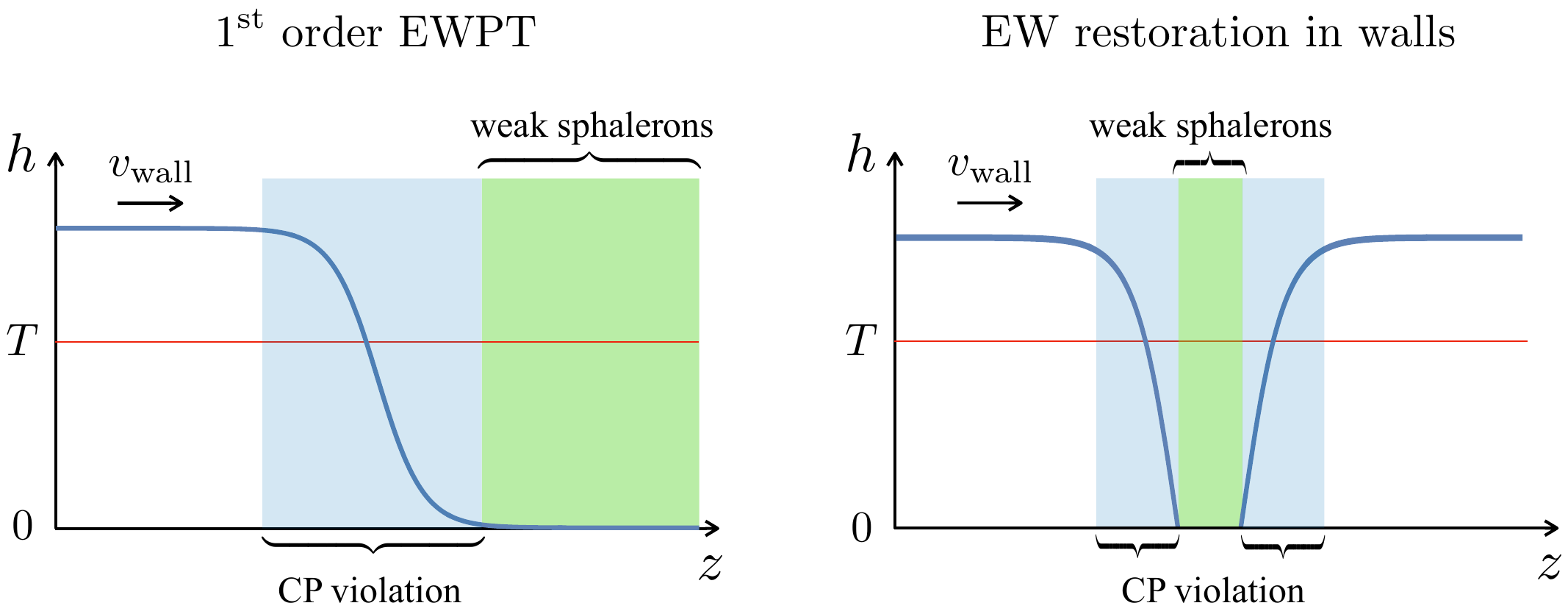}
\caption{Comparison of standard (left panel) and domain-wall assisted (right panel) EWBG.}
\label{fig:sketchintro}
\end{figure}

In the presence of a negative cross-quartic interaction, $-|\lambda_{HS}| |H|^2 S^2$, the effective Higgs mass increases toward the wall core, where $S=0$, potentially inducing a local restoration of the EW symmetry. If such walls persist after electroweak symmetry breaking while featuring electroweak-symmetric cores, the resulting phase structure closely parallels that of a standard first-order electroweak phase transition (EWPT), as illustrated in Fig.~\ref{fig:sketchintro}.

Importantly, such domain walls are not expected to be static. When the characteristic domain size is sufficiently small, their evolution is predominantly driven by the wall tension~$\sigma$, which induces surface contraction. In this regime the walls propagate through the plasma with a typical velocity of order~$0.4$~\cite{Press:1989yh}, while locally deforming toward either of the degenerate minima $\pm v_S$, depending on the local wall curvature. Moreover, a given spatial point may be crossed by a wall multiple times. As a result, the generated baryon asymmetry can either accumulate or partially cancel, depending on the sensitivity of the CP-violating source to the direction of wall motion. In this work we therefore analyze both classes of sources, namely those that are symmetric and those that are antisymmetric under the transformation~$S\to -S$. In the minimal model of Ref.~\cite{Azzola:2024pzq}, the CP-violating sources arise from higher-dimensional operators involving the top quarks, the Higgs field, and the singlet. Such operators can be either even or odd under the $Z_2$ transformation $S\to -S$, thereby realizing precisely the two classes of sources considered below.

A second key factor governing the wall dynamics is the vacuum energy difference~$\Delta V$ between the~$\pm v_S$ domains. A non-vanishing~$\Delta V$ is required to ensure that the walls eventually disappear and do not come to dominate the energy density of the Universe. Depending on the nature of the $Z_2$-breaking phase transition, this vacuum-energy-driven regime may become relevant shortly after wall formation, or only after a period of tension-dominated evolution during which the domains grow to a critical size~$\xi_c$ satisfying~$\Delta V \gtrsim \sigma/\xi_c$. Once this condition is met, the vacuum energy difference controls the subsequent evolution, causing the walls to move preferentially toward the higher-energy vacuum. In this regime neither cancellations nor enhancements arising from multiple wall passages or spatial variations are expected. Furthermore, vacuum-energy domination can lead to an acceleration of the walls to higher velocities. In the reference model of Ref.~\cite{Azzola:2024pzq}, the wall-formation temperature and the onset of vacuum-energy domination are controlled by parameters that are largely independent of those governing wall propagation and baryon-asymmetry generation, and hence these temperatures can be treated as effectively free parameters.

In the present work we remain agnostic about the detailed properties of the $Z_2$-breaking phase transition and instead analyze the resulting baryon asymmetry across the full range of relevant wall velocities, considering both symmetric and antisymmetric sources with respect to the direction of wall propagation. Our main results are formulated for a single wall passage, but they can be straightforwardly generalized to scenarios involving multiple crossings if required. Given the smallness of the baryon asymmetry generated during a single wall passage, the contributions from successive passages can be summed without interference.

Although the singlet-extended SM can realize both standard EWBG during the first-order phase transition, and the domain-wall mediated one, the phenomenological signatures of the two realizations can be markedly different. 
In particular, comparison of DW-assisted EWBG~\cite{Azzola:2024pzq} with standard EWBG in  singlet-extended models~\cite{Espinosa:2011ax,Espinosa:2011eu,Ellis:2022lft,Carena:2019une,Beniwal:2017eik} shows that the DW scenario can favor substantially lower singlet mass scales. 
This opens up a richer phenomenology while remaining compatible with stringent electron EDM bounds~\cite{ACME:2018yjb}, and it may lead to an observable gravitational-wave signal~\cite{Gouttenoire:2025ofv,DEramo:2024lsk,Blasi:2025tmn,Blasi:2023rqi}.

In this paper we present a detailed analysis of baryon asymmetry production in this scenario. 
We assume that the remaining essential ingredient for EWBG---CP violation---arises from a space-dependent phase of the top-quark mass. 
In contrast to Ref.~\cite{Brandenberger:1994mq}, we focus on the regime in which the wall thickness is much larger than the inverse particle momentum. As we show below, this requirement is parametrically close to the condition that the electroweak-symmetric core of the wall be wide enough to contain the weak-sphaleron interaction region, that is necessary for the baryon number production.  
In this thick-wall limit, the effects of a varying complex mass are well described by a semiclassical CP-violating force, which acts with opposite sign on particles and antiparticles. 
The induced local spatial asymmetries in left-handed fermion densities then source baryon number production via weak sphaleron transitions. 
Quantitatively, this dynamics can be captured by a finite set of moments of the Boltzmann equation. 
Moreover, we employ a formulation of the moment expansion that remains applicable at large wall velocities. A special regime of baryon number production in the presence of global charges in the EW-symmetric phase is discussed in a separate publication~\cite{Azzola:2026mbq}.

Recently, a potentially stringent constraint on this class of scenarios has been identified, arising from baryon-asymmetry inhomogeneities~\cite{Bagherian:2025puf,Azatov:2026sdm}. This bound becomes relevant if the typical inter-wall separation grows sufficiently large, eventually reaching a characteristic scale of order~${\cal O}(H^{-1})$~\cite{Press:1989yh,Avelino:2008ve} in the so-called scaling regime, which can be established during the tension-dominated phase of wall evolution. Possible ways to evade the emergence of Hubble-scale separations include additional friction that delays the onset of scaling, or a sufficiently rapid transition to the vacuum-energy-dominated regime, leaving insufficient time for the scaling behavior to develop. The latter option can be straighforwardly realized by a choice of temperatures of the wall network formation and decay. These temperatures are free parameters of the model that do not affect the baryon asymmetry production. 
\

The paper is organized as follows. 
In Section~\ref{s:bap} we collect the elements of the formalism needed to compute the baryon asymmetry in the presence of EW-symmetric domain walls. 
In Section~\ref{sec:toymod} we introduce a simplified model that captures the essential physics and allows us to derive the main qualitative features and parametric scalings of the scenario. 
Section~\ref{sec:bench} presents a numerical study of the full transport network within a generic parametrization of the domain-wall setup. 
In Section~\ref{sec:singlet} we apply our findings to a concrete realization of DW-assisted EWBG in a singlet-extended SM. 
We conclude in Section~\ref{s:conc}.

\section{Transport Framework for Domain-Wall Electroweak Baryogenesis} \label{s:bap}

In this paper we largely follow the approach of Ref.~\cite{Cline:2020jre} to analyze baryon asymmetry production\footnote{Earlier works that laid the foundations of this methodology include Refs.~\cite{Joyce:1994zt, Cline:2000nw, Fromme:2006wx}, while Ref.~\cite{Dorsch:2021ubz} advocates an alternative strategy for solving the transport system.}. This formulation shows a good convergence of the moments expansion at large bubble-wall velocities, which allows us to restrict attention to leading-order effects, as we do in the following.

In this section, we will review the basic logic of baryon asymmetry generation in non-local EWBG scenarios and list the main equations that allow one to quantify it. 
CP-violating (CPV) interactions in the SM plasma can create local excesses of left-handed antiparticles over their respective particles, accompanied by opposite excesses in the right-handed states (chiral asymmetry). The net baryon number density $n_B$ can then be created by the weak sphalerons with a rate $\dot n_B \propto \Gamma_{\text{ws}} \mu_L$, biased towards baryon number production by the chemical potential $\mu_L$ of the left-handed fermions (only quarks are relevant for this paper), defined as

\be
\mu_L = \frac {N_c} 2 \sum \mu_{q_L} ,
\ee
where $\mu_{q_L}$ are chemical potentials of the left-handed quarks per color degree of freedom, counting separately each $SU(2)_L$ component. In the wall frame, the baryon number density change is driven by the equation~\cite{Cline:2000nw}
\be\label{eq:etaprime}
n_B' =  \frac 3 2 \frac{1}{v_{\text{w}}\gamma_{\text{w}}} \Gamma_{\text{ws}} \left(\mu_L T^2 - {\cal A} \, n_B \right),
\ee
where $N_c=3$, and the derivative is taken with respect to the $z$ coordinate directed along the wall propagation. The wall velocity $v_{\text{w}}$ with the corresponding Lorentz factor $\gamma_{\text{w}}$~\cite{Cline:2020jre} arise when transforming from $\dot n_B$ in the plasma frame to the wall frame, where the profiles are time-independent in the assumed stationary wall expansion regime. 
The weak sphaleron rate can be approximated as\footnote{Here we use the numerical pre-factor to match the sphaleron rate at $h=0$~\cite{Bodeker:1999gx} where its precise value is most important, and on top apply the exponential suppression appearing in the broken EW phase. See also Ref.~\cite{Li:2025kyo} for a recent analysis of the sphaleron rate.}
\be
\Gamma_{\text{ws}} \simeq 10^{-6} \, T \, e^{-37 h(z)/T},
\ee
and thus highly suppressed in the regions with a non-zero Higgs vacuum expectation value (VEV) $h$.

The second term on the {\it rhs} of Eq.~\eqref{eq:etaprime} encodes the washout of the baryon asymmetry once the CPV-induced chiral bias $\mu_L$ has relaxed. This occurs, for instance, in the broken phase behind the wall, or in the unbroken region between the two faces of a domain wall, where one eventually approaches an equilibrated excess of left- and right-handed baryons over antibaryons. The resulting baryon chemical potential then biases weak sphaleron transitions in the opposite direction, reducing the net baryon number. Although sphaleron transitions are suppressed in the broken phase, washout can still be appreciable on a Hubble timescale unless the usual condition $h/T\gtrsim 1$ is satisfied.

The factor ${\cal A}$ relates the baryon number density to the chemical-potential bias entering the weak sphaleron rate~\cite{Cline:2000nw} and, in principle, depends on the timescale over which washout is evaluated. We adopt ${\cal A}=15/2$, appropriate for washout on the sphaleron timescale; on a Hubble timescale one finds a slightly smaller value, e.g.\ ${\cal A}=13/2$, which does not  affect our results.

Eq.~\eqref{eq:etaprime} can be integrated to take the form~\cite{Bruggisser:2017lhc,Cline:2020jre} 
\be\label{eq:etagen}
\eta = \frac{135}{4 \pi^2 v_{\text w} \gamma_{\text w} g_* T} \int_{-\infty}^{+\infty} dz \, \Gamma_{\text{ws}} \, \mu_L \, e^{-\frac 3 2 \frac{\cal A}{v_{\text w} \gamma_{\text w}} \int_{-\infty}^z d z_0 \Gamma_{\text{ws}}},
\ee
where $\eta$ is the baryon-to-entropy ratio and $g_*$ is the number of relativistic degrees of freedom. Throughout this paper, $\eta$ denotes the baryon asymmetry produced at a given point by a single wall passage. Residual baryon and lepton asymmetries from earlier passages are negligible compared with the local particle-antiparticle asymmetries generated during a subsequent crossing. Consequently, interference between multiple wall passages is negligible, and the total baryon yield is obtained by summing the contributions from each passage.

The next step is to find the quark chemical potentials $\mu_{q_L}$. To this end one can start from the relativistic Boltzmann equation for the particle distribution functions $f_i$ in the wall frame
\be\label{eq:boltz}
\left\{v_g \partial_z  + F \partial_{k_z} \right\} f_i(z,k)  = {\cal C}[f_j],
\ee
where $v_g$ is the particle velocity, the second term on the {\it lhs} describes the action of an external force, while the collision term on the {\it rhs} describes interactions between different species. The force, including a CPV component, is assumed to be generated by the following background-dependent top quark mass (see Ref.~\cite{Bruggisser:2017lhc} for the case of several quarks with a mass mixing) 
\be
m_t(z) = |m_t|e^{i \theta_t},
\ee
where the spatial variation of the complex phase $\theta_t$ produces CPV, although $|m_t|(z)$ dependence is also important for the exact shape of the CPV source.
One of the ways to derive the force $F$ is to consider a Dirac equation for a quasiparticle in plasma, with a slowly varying Higgs-dependent mass~\cite{Kainulainen:2002th,Fromme:2006wx}. The approximate WKB solution's group velocity is then interpreted as the particle velocity, while its variation allows to infer the classical force $F$.

The WKB derivation of a classical force is valid for slow mass variation, i.e. the mass variation scale $(m_t'/m_t)^{-1}$ has to be greater than the typical inverse particle momentum $1/T$. This is consistent with our mechanism, since the requirement that weak sphalerons fit inside the electroweak-restored regions typically implies a wall width large compared with the inverse temperature, see Fig.~\ref{fig: SSMB}.

After having obtained all the components of the Boltzmann equation~\eqref{eq:boltz}, one can use the following approach for solving it. First, the following ansatz is assumed for the distribution functions~\cite{Fromme:2006wx}
\be\label{eq:fluan}
f_i = \frac{1}{e^{\beta(\gamma_{\text{w}}(E_\text{w} + v_\text{w} p_z) - \mu_i)} \pm 1} + \delta f_i,
\ee
where $\beta = 1/T$, $\pm$ is for fermions/bosons, $\mu_i$ characterizes the deviation of the number density from equilibrium, and $\delta f_i$ parametrizes the remaining deviations, not affecting the particle density, i.e. $\int d^3p \, \delta f = 0$.
Furthermore, instead of trying to solve for the full momentum dependence of $f$, one considers a discrete set of moments. In particular, the first moment of $\delta f$  
\be
u_i \equiv - \frac{3}{2 \pi^3\gamma_{\text{w}} T^2} \int d^3p \, \frac{p_z}{E} \delta f_i
\ee
gives the gross characterization of the imbalance of the particle momentum along or against the wall direction, i.e. the flow of perturbations.
The two sets of equations defining $\mu_i(z)$ and $u_i(z)$ are obtained as moments of Eq.~\eqref{eq:boltz}: first, by integrating Eq.~\eqref{eq:boltz} over $d^3 k$, and second by multiplying with $k_z/E$ and then integrating. Furthermore, we will concentrate on particle-antiparticle asymmetries $(\mu_i - \mu_{\bar i})/2$, $(u_i - u_{\bar i})/2$, referring to them as $\mu_i$ and $u_i$ for simplicity. Expanding in small $\mu$, this results in a system of fluid equations for the asymmetries:
\bea\label{eq:fluid}
-D_1 \,\mu'_i + u'_i + v_{\text{w}} \gamma_{\text{w}} (m^2)' Q_1 \,\mu_i &=& (S^o_{h 1})_i + (\delta C_1)_i,\\
-D_2 \,\mu'_i - v_{\text{w}} \,u'_i + v_{\text{w}} \gamma_{\text{w}} (m^2)' Q_2 \,\mu_i +  (m^2)' \bar R \,u_i &=& (S^o_{h 2})_i + (\delta C_2)_i,
\eea
where the collision terms can be found e.g. in Ref.~\cite{Bruggisser:2017lhc,Cline:2020jre}. Schematically, they have a form
\bea
(\delta C_1)_i 
&=& \left\{ 
\mu_i \sum_{j, \Gamma} \, \Gamma_{i \to j}   
- \sum_{j, \Gamma} \, \Gamma_{j \to i} \,  \mu_j
+ h \, \Gamma_{\text{ss}} \sum_f (\mu_{f_R} - \mu_{f_L})
\right\},
\\
(\delta C_2)_i &=& - \Gamma_{\text{tot}, i} u_i - v_{\text{wall}} \, (\delta C_1)_i,
\eea
where $j$ runs over all particles, $f$ over all fermions;
summation $\sum_{\Gamma} \, \Gamma_{j \to i}$ means summation over all reactions having $j$ in the initial state and $i$ in the final. Negative $\Gamma_{j \to i}$ should be taken when one of $i,j$ is a particle and the other -- antiparticle, to account for a flip of sign of one of the $\mu$ in this case. $h=\pm 1$ for right/left-handed fermions and $h=0$ for bosons. Strong sphalerons occurring with a rate $\Gamma_{\text{ss}}$ tend to relax any excess of $|\mu_L - \mu_R|$.
$\Gamma_{\text{tot}}$ are total elastic rates. For simplicity we suppress the multiplicity factors associated with the number of colors and other possible degeneracies. We adopt numerical expressions for the rates given in Ref.~\cite{Cline:2021dkf}.  

The CP-violating sources are:
\be\label{eq:cpvfull}
(S^o_{h l})_i = - h v_{\text{w}} \gamma_{\text{w}} \left\{ (m_i^2 \theta_i')' Q_{l}^{8o} - (m_i^2)'m_i^2 \theta_i' Q^{9o}_{l} \right\}.
\ee

The coefficient functions $D_i, Q_i, \bar R$ are integrals of the equilibrium distribution functions and their derivatives, and depend on the wall velocity, particle mass, temperature, and spin. They are given in Ref.~\cite{Cline:2020jre} up to an obvious $T$ normalization. 

Further moments can be analyzed to improve precision, and can affect the final asymmetry by an order-few factor~\cite{Kainulainen:2024qpm}. In this work we are restricting ourselves to the first two moments, and hence $\mu$ and $u$ variables, minimally necessary for describing the two key phenomena of non-local EWBG -- local imbalances of particle numbers, and their diffusion.

\section{Analytic Understanding from a Simplified Transport Model}\label{sec:toymod}

The full transport system in Eq.~\eqref{eq:fluid}, which involves all quark and Higgs degrees of freedom and can feature widely separated length scales, is often numerically challenging. We therefore begin with a simplified setup containing only two perturbations and derive analytic expressions for their behaviour and for the resulting baryon asymmetry in several relevant limiting regimes. This reduction serves two purposes: it builds physical intuition for the mechanism of asymmetry generation and, as we show later, yields in certain cases a rather accurate approximation to the baryon asymmetry obtained from the complete system.
  
To this end we will use a simplified version of the transport network~\eqref{eq:fluid} involving only two fluctuations, namely the chemical potential $\mu_{t_L}$ and the corresponding velocity perturbation $u_{t_L}$ that we refer to as $\mu$ and $u$ for simplicity: 
\bea\label{eq:toytransport}
\left(\begin{matrix} 
|D_1| & 1 \\ 
D_2 & v_{\text w}
\end{matrix} \right)
\left(\begin{matrix} 
\mu' \\ 
u'
\end{matrix} \right)
- 
\left(\begin{matrix} 
\Gamma_\mu  & 0 \\ 
v_{\text{w}} \Gamma_\mu & \Gamma_u
\end{matrix} \right)
\left(\begin{matrix} 
\mu \\ 
u
\end{matrix} \right)
=
\left(\begin{matrix} 
S_1 \\ 
-S_2
\end{matrix} \right).
\eea
The first term describes the flow of perturbations from/to the neighboring regions, the second -- their decay, and {\it rhs} contains their source. 
Here the negative coefficient function $D_1$ was written as $-|D_1|$ to show the signs explicitly. We have also assumed the term proportional to $(m^2)'$ in the free part of Eqs.~\eqref{eq:fluid} to be subdominant and dropped it. Furthermore, we set the top mass to zero in the free part, i.e. when computing $D_{1,2}$ and $\Gamma_{\mu}$.  
The typical values of coefficient functions and sources are
\be\label{eq:coefsize}
|D_1| \simeq v_{\text {w}}
\,,\;\; 
D_2 \lesssim 1
\,,\;\;
|S_1/S_2| \simeq v_{\text {w}}\,,
\ee
while the decay rates are typically of order 
\be
\Gamma_{\mu,u} \sim 10^{-1\ldots-2} T.
\ee

First, let us derive the most naive estimate of the baryon asymmetry produced, without accounting for any further suppression factors, that will be analyzed later on. 
The expression Eq.~\eqref{eq:etagen} can be approximated as
\bea
\eta 
\simeq  
C_\eta \int_{- \infty}^{+\infty} dz  \, e^{-37h/T} \, \mu, 
\quad\text{with} \quad
C_\eta = 10^{-6}\frac{135 N_c}{8 \pi^2 v_{\text w} \gamma_{\text w} g_*}.
\eea
Assuming the baryon density is produced from the chiral asymmetry over the wall width $l_{\text{wall}}$ we find
\be
\eta 
\,\sim\,  
C_\eta \, l_{\text{wall}} \,\langle \mu \rangle.
\ee
The typical chiral asymmetry $\langle \mu \rangle$ can be estimated using dimensional analysis 
\be
\langle \mu \rangle 
\,\sim\,
\int dz' S(z')
\,\sim\,
l_{\text{source}} \langle S \rangle,
\ee
where $l_{\text{source}}$ is the spatial extent of the CPV source~\eqref{eq:cpvfull},
that has a typical magnitude 
\be
\langle S \rangle 
\,\sim\, 
v_{\text w} \gamma_{\text w} \frac{\Delta\theta}{l_{\text{source}}^2} \langle m_t^2 Q^{8o}_{2} \rangle 
\,\sim\, 
5 \cdot 10^{-2} \, v_{\text w} \gamma_{\text w} \frac{\Delta\theta}{l_{\text{source}}^2},
\ee
with $\Delta\theta$ being the maximal variation of the top quark phase across the wall and we used a maximal numerical value of the $m_t^2 Q_{8o2}$ coefficient entering the source. Taking $g_* = 107$ (valid for $T\sim100$~GeV), we obtain the baryon asymmetry estimate
\bea\label{eq:etanaive}
\eta_{\text{naive}}
&\sim& 5 \cdot 10^{-9} \,\Delta\theta \frac{l_{\text{wall}}}{l_{\text{source}}},
\eea
while the required number is $\eta \simeq 10^{-10}$. This naive estimate neglects the interplay between different regions of the wall and implicitly assumes that only two length scales are relevant. In practice, however, the perturbation decay length $l_{\Gamma}$ plays a crucial role in determining the resulting baryon asymmetry. 
Typically, \(l_\Gamma \sim \Gamma^{-1}\), where \(\Gamma\) denotes the relevant interaction rates in the plasma, so this scale is typically determined by the temperature and the SM couplings. By contrast, in DW-assisted baryogenesis~\cite{Azzola:2024pzq}, both the wall width ($l_{\text{wall}}$) and the CP violating source size ($l_{\text{source}}$) are set by the inverse singlet mass and may therefore vary over several orders of magnitude, with \(l_{\text{source}} < l_{\text{wall}}\) in typical realizations. As a result, essentially any hierarchy among the three length scales, consistent with \(l_{\text{source}} < l_{\text{wall}}\), can be realized. In the following, we study how the naive estimate~\eqref{eq:etanaive} is modified as this hierarchy is varied, and how the result depends on the structure of the CP-violating source.

\begin{figure}[t]
\center
\includegraphics[width=0.5 \textwidth]{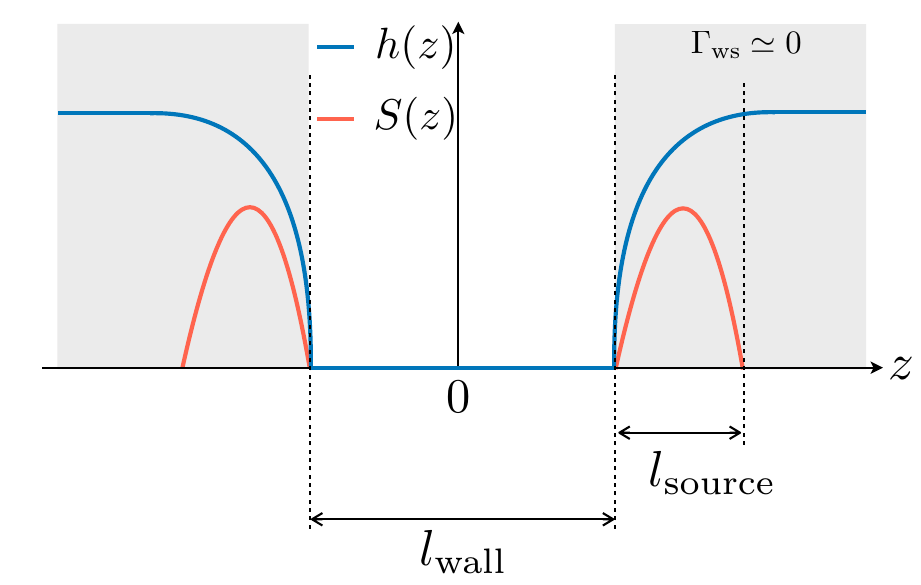}
\caption{Schematic representation of the domain wall with vanishing Higgs VEV inside. The CPV source is assumed to be associated with the Higgs VEV variation, hence there are two regions with non-vanishing source --- at the front and the rear parts of the wall. The Higgs VEV variation length determines the characteristic length of the source.}
\label{fig:scales}
\end{figure}

\subsection{Wide wall, narrow source ($l_{\text{source}} \ll l_\Gamma \ll l_{\text{wall}}$)}
\label{ss: Chiral Asymmetry}

Let us first analyze the simplest situation $l_{\text{source}} \ll l_\Gamma \ll l_{\text{wall}}$ when the sources can be approximated as point-like, $S_i(z) = \hat s_i \delta(z-z_s)$, with the coefficients $\hat s_i$ defined from matching to an actual extended source
\bea
&\int dz \, S_i(z) 
= \int dz \, \hat s_i \, \delta(z-z_s)& \\
&\Rightarrow \hat s_i = l_{\text{source}} \langle S_i \rangle,&  \label{eq:sourcematch}
\eea
with $\langle S_i \rangle$ being averaged source values. For now we consider a single point-like source at $z=z_s$, without its counterpart on the opposite side of the wall, see Fig.~\ref{fig:scales}.

\begin{figure}[t]
\center
\includegraphics[width=0.49 \textwidth]{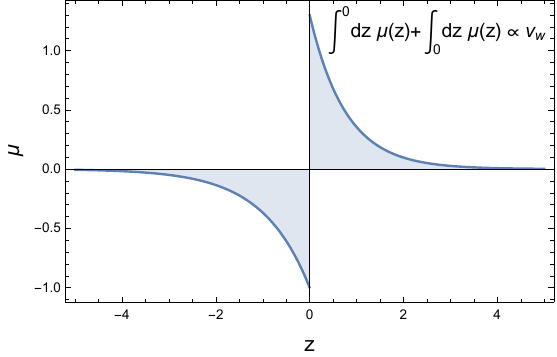}
\caption{Schematic behaviour of the chiral asymmetry $\mu$ (in arbitrary units) produced by a point-like source at $z=0$.}
\label{fig:sketchmu}
\end{figure}

Before solving the system~\eqref{eq:toytransport} explicitly, it is useful to anticipate the qualitative structure of its solution. A point-like source generates two decaying perturbations: one in front of the wall, corresponding to diffusion ahead of the source, and one behind it, corresponding to the trail left by the passing wall. In both regions, interactions drive the system back to equilibrium, leading to exponential damping of perturbations away from the source. This structure is illustrated in Fig.~\ref{fig:sketchmu}.

We now turn to a detailed derivation of the solution. We first solve the homogeneous part of the system~\eqref{eq:toytransport} and then incorporate the source term.
The solutions of the homogeneous system read:
\be\label{eq:muupm}
\left(\begin{matrix} 
\mu \\ 
u
\end{matrix} \right)_\pm
=
 c_\pm e^{\lambda_\pm (z-z_s)} 
v_\pm,
\ee
where $(\lambda_\pm, v_\pm)$ are pairs of eigenvalues and eigenvectors of the operator
\bea\label{eq:toytransport2}
\left(\begin{matrix} 
|D_1| & 1 \\ 
D_2 & v_{\text w}
\end{matrix} \right)^{-1}
\left(\begin{matrix} 
\Gamma_\mu  & 0 \\ 
v_{\text{w}} \Gamma_\mu & \Gamma_u
\end{matrix} \right),
\eea
and define respectively the trailing and diffusing solutions mentioned above. 
At small velocities
\bea \label{eq:lambdaapprox}
\lambda_{\pm} 
&\simeq& 
\pm \sqrt{\frac{\Gamma_\mu \Gamma_u}{D_2}} - \epsilon \frac{|D_1|}{2 D_2} \Gamma_u.
\eea

Here and in the following we formally write powers of $\epsilon=1$ in front of the terms that are suppressed by respective powers of $v_{\text w}$ (see Eq.~\eqref{eq:coefsize}). In this paper we will be focusing on wall velocities in the range $v_{\text{w}} \sim 0.1 \ldots 1$. 
According to Eq.~\eqref{eq:lambdaapprox} there is one growing and one decaying solution, which remains true also for large velocities, see Ref.~\cite{Cline:2020jre}. The form of the full solution (basically, the Green's function normalized with $\hat s_i$)
\be \label{eq:rtoy}
\mu(z,z_s) = 
\theta(z_s-z) \mu_+  
+
\theta(z-z_s) \mu_- 
\ee
is fixed by demanding the perturbations to decay away from the source located at $z=z_s$.
Importantly, at $v_{\text{w}}\to 1$ one finds that $\lambda_- \to -\infty$, which means that the extent of the forward ($z>z_s$) tail of the $\mu$ distribution~\eqref{eq:rtoy} shrinks to zero~\cite{Cline:2020jre}. This is expected as the perturbations cannot travel faster than the wall in this limit. 

The coefficients $c_\pm$ in Eq.~\eqref{eq:muupm} can be fixed from the discontinuity induced by the source:
\be
\label{cpcm fix}
\left(\begin{matrix} 
|D_1| & 1 \\ 
D_2 & v_{\text w}
\end{matrix} \right)
(c_- v_- - c_+ v_+) 
= 
\left(\begin{matrix} 
\hat s_1 \\ 
-\hat s_2
\end{matrix} \right),
\ee
so that the two tails of the full solution read, in small-$v_{\text w}$ approximation
\be \label{eq:muampapprox}
\mu_\pm = \left( \pm \frac{\hat s_2}{2 D_2} 
- \epsilon \, \frac{2 D_2 \hat s_1 + |D_1| \hat s_2}{4 D_2^{3/2}} \sqrt{\frac{\Gamma_u}{\Gamma_\mu}} \right) e^{\lambda_\pm (z-z_s)},
\ee

From the expressions for the $\mu$ fluctuations amplitude~\eqref{eq:muampapprox} it is clear that at low wall velocities it simply flips sign across the source, with an almost unchanged absolute value. Furthermore, it decays with a similar rate in front and behind the wall, see Eq.~\eqref{eq:lambdaapprox}.  As a result, the integrated chiral asymmetry cancels out up to corrections $\propto v_{\text w}$:
\be\label{eq:pointmuint}
\int_{-\infty}^{+\infty}  dz \, \mu(z) 
= 
 - \epsilon \frac{\hat s_1}{\Gamma_\mu},
\ee
which is illustrated in Fig.~\ref{fig:sketchmu}. 
Such a situation is expected since at low wall speeds the fluctuations carried by particles with speeds of order-one would distribute almost symmetrically in both directions, up to an overall sign of $\mu$.

\vspace{0.5cm}

We now modify the source to better capture the structure of an actual domain-wall configuration. The domain walls of interest exhibit a decreasing Higgs VEV at the leading edge and an increasing VEV at the trailing edge, as illustrated in Fig.~\ref{fig:scales}. This structure effectively gives rise to two spatially separated sources of CP violation. To capture this behavior in a minimal way, we model the system using two delta-function sources, arranged either symmetrically or antisymmetrically with respect to the wall center. Thereby they can mimic for example the CPV sources of the model presented in Ref.~\cite{Azzola:2024pzq} (see Eq.~\eqref{eq: CPV}):
\bea\label{eq:twopointsources}
S_{\text{even}\,i}(z) 
&=& 
\hat s_i  
\left(\delta(z+l_{\text w}/2) + \delta(z-l_{\text w}/2)\right),
\\
S_{\text{odd}\,i} (z)
&=& 
\hat s_i \left(\delta(z+l_{\text w}/2) - \delta(z-l_{\text w}/2)\right).
\eea

\begin{figure}[t]
\center
\includegraphics[width=0.48 \textwidth]{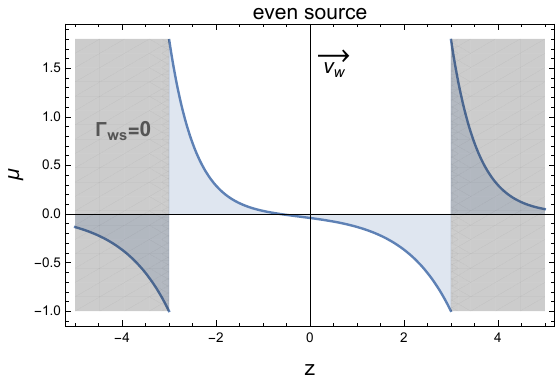}
\hspace{0.1cm}
\includegraphics[width=0.48 \textwidth]{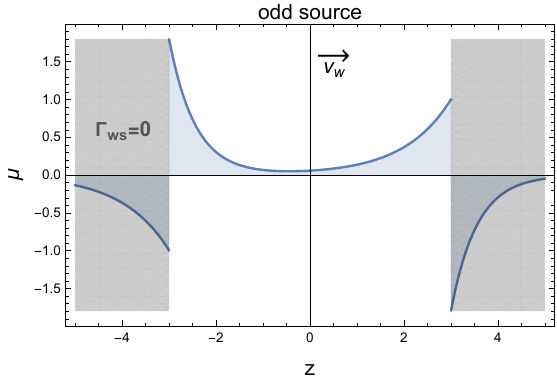}
\caption{Schematic behaviour of the chiral asymmetry $\mu$ (in arbitrary units) produced by an even and an odd combination of point-like sources. Weak sphalerons are suppressed in gray regions.}
\label{fig:sketchmuevenodd}
\end{figure}

Furthermore, we assume that electroweak sphalerons are completely switched off for $|z|>l_{\text{wall}}/2$ and are unsuppressed within the wall. As a result, sphalerons act only on the trailing tail of the perturbations at the front edge of the domain wall, and only on the leading tail at the rear edge. This situation is illustrated in Fig.~\ref{fig:sketchmuevenodd}.

Consequently, for an even source the effective chemical-potential profile relevant for baryon production is essentially the same as in the single delta-function case, up to an interchange of the two tails, cf. Fig.~\ref{fig:sketchmu} and the middle area of the left panel of Fig.~\ref{fig:sketchmuevenodd}. At small $v_{\text{w}}$, the net contribution again suffers from a cancellation between two regions of opposite sign. For an odd source, by contrast, one of the $\mu$ profiles flips sign, thereby removing this cancellation, as shown in the right panel of Fig.~\ref{fig:sketchmuevenodd}.

The baryon asymmetry is simply given by
\bea 
\eta \simeq C_\eta \int_{-l_{\text{wall}}/2}^{l_{\text{wall}}/2} dz \, \mu(z) 
&=& 
C_\eta \left\{
P_S \frac{c_+ v_+}{\lambda_+} \left(1 - e^{- \lambda_+ l_\text{wall}}\right)-
 \frac{c_- v_-}{\lambda_-} \left(1 - e^{\lambda_- l_\text{wall}}\right)
\right\} \quad \quad \label{eq:muint1}\\
&\underset{l_{\text{wall}}\gg 1/\lambda_\pm}{\simeq} &
C_\eta \left\{
P_{S} \frac{c_+ v_+}{\lambda_+} - 
 \frac{c_- v_-}{\lambda_-}
\right\}, 
\eea
where $P_{S}=\pm1$ for the even and the odd source respectively, and $c_\pm v_\pm$ are the solutions for the single point source defined in Eq.~\eqref{cpcm fix}.
We then find, explicitly:
\bea 
\eta^{(l_{\text{source}} \ll l_\Gamma \ll l_{\text{wall}})}_{\text{\bf even}} 
&=&
C_\eta \,\epsilon\, \frac{- \hat s_1}{\Gamma_\mu}, \label{eq:eta11}
\\
\eta^{(l_{\text{source}} \ll l_\Gamma \ll l_{\text{wall}})}_{\text{\bf odd}} 
&\underset{v_{\text{w}\ll 1}}{\simeq} & 
C_\eta \frac{-\hat s_2}{\sqrt{D_2 \Gamma_u \Gamma_\mu}}, \label{eq:eta12}
\\
\eta^{(l_{\text{source}} \ll l_\Gamma \ll l_{\text{wall}})}_{\text{\bf odd}} 
&\underset{v_{\text{w}\to 1}}{\simeq} & 
C_\eta \left( \frac{\hat  s_1}{\Gamma_{\mu}} - 2\frac{\hat  s_1 + \hat s_2}{\Gamma_u}  \right),
\label{eq:eta13}
\eea
where the expression for the even case is exact, while for the odd case we showed small- and large-$v_{\text{w}}$ expansion\footnote{In the assumed massless limit 
$D_2 = (-2 v + 4 v^3 + 2 (v^2-1)^2 \tanh^{-1}[v])/(2 v^3)$.}.

Aside from reducing the amount of cancellation, too large wall speeds can suppress the CP-violating source. By examining the $Q_{8o1,8o2,9o1,9o2}$ coefficient functions entering the source in Eq.~\eqref{eq:cpvfull} one can see that they drop as $1/\gamma^p$ with $p \sim 1.5$, determined numerically.

Overall, combining the suppression effects discussed in this section, we obtain the following parametric estimates of the baryon asymmetry:
\bea\label{eta:pointsource}
\eta^{(l_{\text{source}} \ll l_\Gamma \ll l_{\text{wall}})}
&\sim& \eta_{\text{naive}} \times
\begin{cases}
({v_{\text{w}}}/{\gamma^p}) \, ({l_\Gamma}/{l_{\text{wall}}}) & \text{even $S$}\\
({1}/{\gamma^p}) \, ({l_\Gamma}/{l_{\text{wall}}}) &\text{odd $S$}
\end{cases}
\eea
where the naive estimate $\eta_{\text{naive}}$ was defined in Eq.~\eqref{eq:etanaive}.
The ${l_\Gamma}/{l_{\text{wall}}}$ ratio appears since the chiral asymmetry only spreads over the distance $l_\Gamma \sim 1/\Gamma_{\mu, u}$ (present in each of Eqs.~\eqref{eq:eta11}, \eqref{eq:eta12}, \eqref{eq:eta13}), rather than the full wall width, as was assumed for the naive estimate.

Finally, one can note that there is yet another source of baryon asymmetry suppression for ultra-relativistic walls. The width of such walls in plasma frame contracts, leaving less space for the sphaleron field configurations to appear. When the wall size drops below $\sim 1/\alpha_w T$ the sphalerons are expected to be exponentially less frequent~\cite{Arnold:1987mh}.

\subsection{Narrow wall and source ($l_{\text{source}} \ll l_{\text{wall}} \ll l_\Gamma$)}
\label{sec:wall_thick}

We now analyze the case of a finite wall width: $l_{\text{wall}} \ll l_\Gamma$. This also implies $l_\text{source} \ll l_\Gamma$, hence we will again assume the same even and odd point-like sources of Eq.~\eqref{eq:twopointsources}.

The chiral asymmetry integrated over the active sphaleron region is given by the small-$l_{\text{wall}}$ limit of Eq.~\eqref{eq:muint1}: 
\bea 
\eta \simeq C_\eta \int_{-l_{\text{wall}}/2}^{l_{\text{wall}}/2} dz \, \mu(z)
&\underset{l_{\text{wall}}\ll 1/\lambda_\pm}{\simeq} &
C_\eta \,l_{\text{wall}} \left\{
P_S\, c_+ v_+ + c_- v_-
\right\}.
\eea
Explicitly, one finds
\bea 
\eta^{(l_{\text{wall}}\ll l_\Gamma)}_{\text{\bf even}} 
&\underset{v_{\text{w}\ll 1}}{\simeq} &
- C_\eta \, \epsilon \, \frac{2 D_2 \hat s_1 + |D_1| \hat s_2}{2 D_2^{3/2}}  \sqrt{\frac{\Gamma_u}{\Gamma_\mu}} 
\,l_{\text{wall}},
\\
\eta^{(l_{\text{wall}}\ll l_\Gamma)}_{\text{\bf odd}} 
&=& 
C_\eta\, \frac{\hat s_2 + v_{\text{w}} \hat s_1}{|D_1| v_{\text{w}} - D_2}
\,l_{\text{wall}}.
\eea
We do not present the $v_{\mathrm{w}}\to 1$ limit for the even source here, since this regime requires a careful treatment of the condition $l_{\text{wall}} < 1/|\lambda_-|$. This condition becomes potentially violated since $1/|\lambda_-|\to 0$ as $v_{\mathrm{w}}\to 1$.

Comparing this to the baryon asymmetry estimate derived in the beginning of this section we see that the weak sphalerons convert $\mu$ into $\eta$ over the entire wall width, hence the original naive estimate actually holds. The previously found additional $v_{\text{w}}$-suppression for the even source however persists. Overall, for such thin walls we get the following parametric estimate
\bea\label{eta:thinwall}
\eta^{(l_{\text{wall}}\ll l_\Gamma)}
&\sim& \eta_{\text{naive}} \times
\begin{cases}
{v_{\text{w}}}/{\gamma^p} & \text{even $S$}\\
{1}/{\gamma^p} &\text{odd $S$}
\end{cases}
\eea

\subsection{Wide wall and source ($l_\Gamma \ll l_{\text{source}} \ll l_{\text{wall}}$)}

\begin{figure}[t]
\center
\includegraphics[width=0.45 \textwidth]{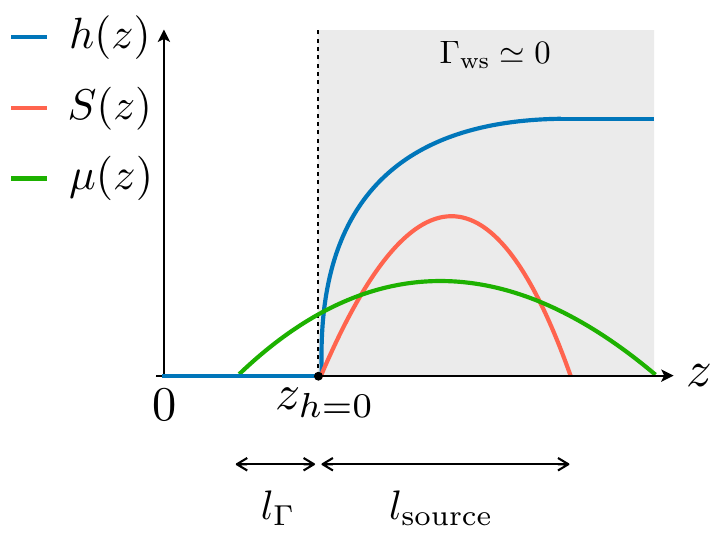}
\caption{Chiral asymmetry produced by an extended source at the front side of the wall. Sphalerons are inactive in the region where the source is localized, hence only the tail of the $\mu$ distribution leaking into the $l_\Gamma$-deep layer of $h=0$ region is relevant for baryon asymmetry production.}
\label{fig:ext_source}
\end{figure}

Let us now analyse the regime $l_\Gamma \ll l_{\text{source}} \ll l_{\text{wall}}$, where the source size cannot be neglected. 
We start by considering the forward side of the wall, schematically depicted in Fig.~\ref{fig:ext_source}. The baryon asymmetry is mostly produced in the region with a vanishing source but unsuppressed EW sphalerons, at $z<z_{h=0}$:
\be
\eta \simeq C_\eta \int_{z_{h=0}-l_\Gamma}^{z_{h=0}} dz\  \mu(z), 
\ee
where the $\mu$ distribution decays in the absence of the source, according to Eq.~\eqref{eq:rtoy}:
\be
\mu(z\leq z_{h=0}) = \mu(z_{h=0}) \, e^{\lambda_{+} (z-z_{h=0})}.
\ee
We then find a $l_\Gamma = 1/\lambda_+$ suppression for the baryon asymmetry:
\be\label{eq:eta31}
\eta \simeq C_\eta \, \mu(z_{h=0})/\lambda_{+}.
\ee 
On the other hand, the chiral asymmetry produced at the $z_{h=0}$ boundary is accumulated from the sources at $z>z_{h=0}$, and again only over a distance $\sim 1/\lambda_{+}$, which is the maximal distance that perturbations propagate. Specifically, we obtain
\bea
\mu(z_{h=0}) &=& \int_{z_{h=0}}^{\infty} dz' \, \mu(z_{h=0},z') \nonumber \\
&=& \int_{z_{h=0}}^{\infty} dz' \,  \,e^{\lambda_+ (z_{h=0}-z')}  \tilde c_+ (z') v_+ \nonumber \\
&\simeq& \tilde c_+ (z_{h=0}) v_+ / \lambda_+.\label{eq:mueintxt}
\eea
In the second step we used the extended source generalization of the point-like source solution~\eqref{eq:rtoy}. In particular, $\tilde c_\pm$ are now determined by analogy with Eq.~\eqref{cpcm fix}, with $S_i(z)$ replacing  $\hat s_i$. In the final step we assumed that $\tilde c_+$ doesn't vary much over the decay length $1/\lambda_+$. Overall, this brings a factor $l_\Gamma /l_\text{source}$ to $\eta$, due to the fact that only the $l_\Gamma$-wide region of the source right in front of the wall can contribute to the asymmetry. 

Including now both sides of the domain wall, with a symmetric or antisymmetric source, we find
\bea 
\eta^{(l_\Gamma \ll l_{\text{source}})} 
&\simeq&  
C_\eta
\left\{
P_S \frac{\tilde c_+ v_+}{\lambda_+^2} +
\frac{\tilde c_- v_-}{\lambda_-^2}
\right\},
\eea
or, explicitly
\bea 
\eta_{\text{\bf even}}^{(l_\Gamma \ll l_{\text{source}})} 
&\underset{v_{\text{w}\ll 1}}{\simeq}&
C_\eta \, \epsilon \, \frac{|D_1| S_2 - 2 D_2 S_1}{2 \Gamma_u^{1/2} \Gamma_\mu^{3/2} D_2^{1/2}}, 
\\
\eta_{\text{\bf even}}^{(l_\Gamma \ll l_{\text{source}})} 
&\underset{v_{\text{w}\to 1}}{\simeq}&
C_\eta \frac{\Gamma_\mu(S_1+S_2)  - S_1 \Gamma_u}{\Gamma_u \Gamma_\mu^2},
\\
\eta^{(l_\Gamma \ll l_{\text{source}})}_{\text{\bf odd}} 
&=& 
C_\eta \frac{|D_1| S_1 \Gamma_u - \Gamma_\mu (S_2 + v_{\text w}S_1)}{\Gamma_u \Gamma_\mu^2},
\eea
where $S_1,S_2$ are evaluated at $z = -|z_{h=0}|$~\footnote{More precisely, these are average source values in the $\Delta z = 1/\lambda_\pm$ layer next to the active EW sphaleron region.}.

Parametrically, the baryon asymmetry now features $l_\Gamma^2$ suppression originating from $1/\lambda$ factors in Eqs.~\eqref{eq:eta31} and~\eqref{eq:mueintxt}:
\bea\label{eta:extsource}
\eta^{(l_\Gamma \ll l_{\text{source}})}
&\sim& \eta_{\text{naive}} \times
\begin{cases}
({v_{\text{w}}}/{\gamma^p})
\,({l_\Gamma^2}/{l_{\text{wall}} l_{\text{source}}}) & \text{even $S$}\\
({1}/{\gamma^p}) \, ({l_\Gamma^2}/{l_{\text{wall}} l_{\text{source}}}) &\text{odd $S$}
\end{cases}
\eea

\begin{table}[]
    \centering
    \begin{tabular}{|c|c|c|}
        \hline
        \textbf{Regime} & \textbf{Even source} & \textbf{Odd source}\\
        \hline 
        \hline
        Wide wall, narrow source & $({v_{\text{w}}}/{\gamma^p}) \, ({l_\Gamma}/{l_{\text{source}}})$ & $({1}/{\gamma^p}) \, ({l_\Gamma}/{l_{\text{source}}})$ \\
        \hline
        Narrow wall and source & $({v_{\text{w}}}/{\gamma^p})({l_\text{wall}}/{l_{\text{source}}})$ & $({1}/{\gamma^p})({l_\text{wall}}/{l_{\text{source}}})$ \\
        \hline
        Wide wall and source & $({v_{\text{w}}}/{\gamma^p})
\,({l_\Gamma}/{l_{\text{source}}})^2$ & $({1}/{\gamma^p}) \, ({l_\Gamma}/{l_{\text{source}}})^2$ \\
        \hline
        
    \end{tabular}
    \caption{Parametric dependence of the baryon asymmetry in different regimes. The magnitude of $\eta$ is given by $ 5 \cdot10^{-9}\Delta\theta$ times the factor displayed in the table. $p \sim 1.5$ is determined numerically.}
    \label{tab:summary regimes}
\end{table}

In Table~\ref{tab:summary regimes}, we summarize the parametric dependence of $\eta$ on the physical length scales in the three different regimes.
\section{Full Transport Analysis in a Generic Domain-Wall Setup} 
\label{sec:bench}

In this section we extend the analytical framework of Section~\ref{sec:toymod} to a more realistic setup, specified below. We first enumerate the physical effects included in the full diffusion system that are absent in the toy model. We then compare numerical solutions to the analytical estimates in order to quantify the resulting corrections and to assess the range of validity of the simplified treatment. Finally, we study the temperature dependence of the baryon asymmetry.

We begin by defining the benchmark model used throughout this section. It is characterized by two length scales, $l_\text{wall}$ and $l_\text{source}$, which set the depth of the EW-symmetric region, and the distances over which the Higgs mass and the top-quark phase vary by order one, respectively. The model is specified by the wall profile
\bea
\label{eq:benchmark h}
h^2(z)=F^2T^2\left[1-\kappa\left(1-\tanh^2\left(\frac{z}{l_\text{source}}\right)\right)\right],
\eea
whenever the right-hand side is positive, and $h(z)=0$ otherwise. Here
$$
\kappa=\frac{1}{1-\tanh^2 \left(\frac{l_\text{wall}}{2l_\text{source}}\right)} 
\simeq \frac 1 4 \exp \left(\frac{l_\text{wall}}{l_\text{source}} \right),
$$
where in the last step we showed a large-$l_\text{wall}/l_\text{source}$ asymptote. Although the expression in Eq.~\eqref{eq:benchmark h} may look involved, one can easily check that $h$, being a function of $\tanh(z/l_\text{source})$ has a characteristic decay length $l_\text{source}$. On the other hand, one can also verify that the Higgs VEV vanishes at $z= \pm l_\text{wall}/2$, as required.

As discussed in Section~\ref{sec:singlet}, the $\kappa$ factor can be related to the tuning of the Higgs mass in the singlet-extended SM.

The parameter $F$ defined as 
\be
F=h(\pm\infty)/T
\ee
is the analogue of the strength of a first-order phase transition.

\begin{figure}
\center
\includegraphics[width=0.47 \textwidth]{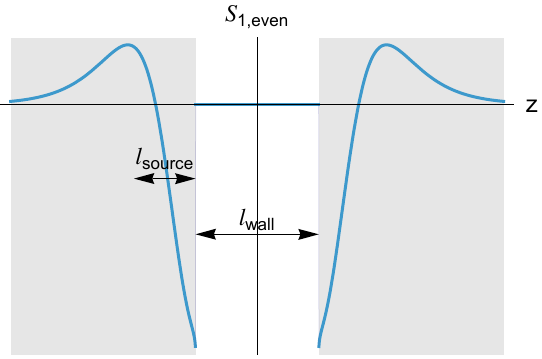}
\hspace{0.3cm}
\includegraphics[width=0.47 \textwidth]{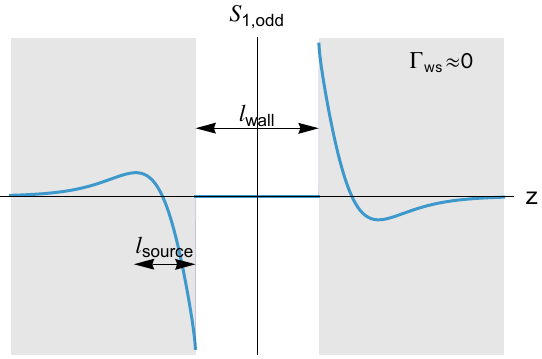}
\caption{Typical source profile for the benchmark model, for $n_\text{CP} = 2$ and $n_\text{CP} = 1$ on the left and right respectively.}
\label{fig:benchmark source}
\end{figure}

The top-quark mass is taken to be
\bea
 \label{eq:benchmark t}
    |m_t(z)| &=& \frac{y_t}{\sqrt{2}} h(z), \\ 
    \arg[m_t(z)] &=& \arctan\left[\left(\frac{\Delta\theta}{2}\right)^{n_\text{CP}} \left(  \tanh\frac{z + l_\text{wall}/2}{l_\text{source}} + \tanh \frac{z - l_\text{wall}/2}{l_\text{source}} \right)^{n_\text{CP}}\right],
\eea
where $n_\text{CP}=1\ (2)$ for an odd (even) source, respectively. See Figure~\ref{fig:benchmark source} for a representative example of the resulting profile. Unless stated otherwise, we fix $\Delta\theta=1$ and $F=1.5$. Since $\eta_B$ scales approximately linearly with $\Delta\theta$, results for other phase variations can be obtained by simple rescaling. A discussion of the effect of $F$ on the produced asymmetry can be found in Section~\ref{s:tedep}. 

\subsection{Physical effects beyond the simplified model} \label{ss:physical effects beyond toy}

We now discuss the effects that are not captured by the simplified model of the previous section, and discuss their impact on the baryon asymmetry.

\paragraph{Number of species.}
Computing $\eta_B$ in full generality requires solving the coupled diffusion equations for all relevant Standard Model species. In practice, however, it is often sufficient to work with a reduced network that retains only the degrees of freedom that dominantly participate in sourcing and transporting the chiral asymmetry. Since the CP-violating source acts directly only on the top sector, one expects the coupled system for $\mu_{t_L}$ and $\mu_{t_R}$ to provide the leading contribution to the left-handed chiral bias $\mu_L$. Accordingly, a network consisting solely of left- and right-handed top quarks often yields a rather accurate approximation to the full solution.

A further simplification follows from the observation that the CP-violating source acts with equal magnitude and opposite sign on $t_L$ and $t_R$, and that the two chiralities share part of their interactions. This suggests that, to a good approximation, $\mu_{t_L}\simeq -\mu_{t_R}$, with an analogous relation holding for the corresponding velocity perturbations. One may then solve a reduced system for the left-handed top alone, using $\mu_{t_R}=-\mu_{t_L}$ to eliminate $\mu_{t_R}$ from the collision terms in the $t_L$ transport equation.

\begin{figure}[]
\center
\includegraphics[width=0.8\textwidth]{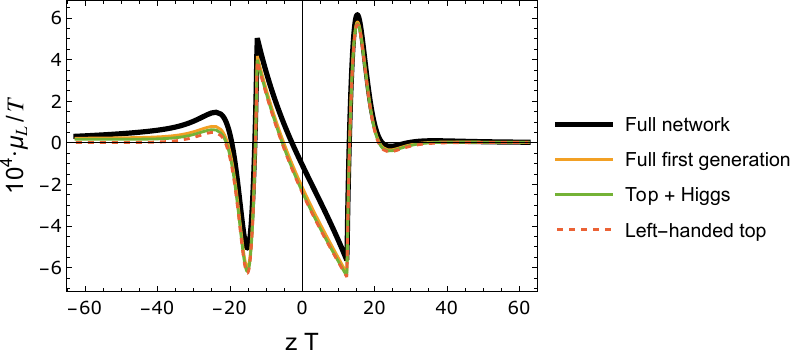}
\caption{Comparison of $\mu_L$ profile for different number of species included in the transport network, see text for details. Parameters set as: $l_\text{wall} = 25$ and $l_\text{source} = 5$ and $v_w = 0.5$, and an even source is used.}
\label{fig: species plot}
\end{figure}

Numerically, we find that reducing the number of species included in the transport network has no significant impact on the resulting baryon asymmetry. To verify this explicitly, we computed $\eta_B$ using the following four networks:
\begin{enumerate}
\item
With 9 species including all quarks in both chiralities and the Higgs field. Following Ref.~\cite{Bruggisser:2017lhc}, we describe the two lightest generations of quarks as the charm-strange pair, with double degrees of freedom. 
\item
With the light quarks dropped.
\item
A network including only top quarks and the Higgs, as motivated above.
\item
A single top-left species, accounting for the right-handed partner via $\mu_{t_R} \approx - \mu_{t_L}$. 
\end{enumerate}

Fig.~\ref{fig: species plot} shows $\mu_L$ profiles in different approximations. The difference in the total baryon asymmetry is always within $10\%$, thus motivating the simplified model for a quantitative exploration of the parameter space.

\paragraph{Finite source size.}
The approximations derived in Section~\ref{sec:toymod} are naturally reliable in the two opposite limits $l_{\rm source}\ll l_\Gamma$ and $l_{\rm source}\gg l_\Gamma$. In both regimes, the strong hierarchy between the source width and the diffusion length implies that the perturbations cannot fully resolve the spatial structure of the source.

We now turn to the intermediate regime in which $l_\Gamma$ becomes comparable to the other two length scales. When $l_{\rm source}\sim l_\Gamma$, the perturbations can begin to resolve the source profile, which in our benchmark model exhibits a two-peak structure with opposite sign on each side of the wall (cf.\ Fig.~\ref{fig:benchmark source}). As a result, depending on the ratio $l_\Gamma/l_{\rm source}$, the perturbations may either experience an effectively averaged and thus suppressed source, or become predominantly sensitive to the peak closest to the electroweak-restored region, leading to an enhancement. In this regime, finite-mass effects can also become important. We therefore first discuss the mass effects before comparing results across different approximations.

\paragraph{Finite top mass.}
A further approximation made in the analytic treatment of Sec.~\ref{sec:toymod} was to treat the left-handed quark as effectively massless when constructing the homogeneous solutions. While this is exact in the electroweak-symmetric phase, top quarks propagate as massive excitations outside the wall. Since diffusion lengths are typically shorter for heavier species, including the finite top mass reduces the sensitivity to source regions that are not directly adjacent to the electroweak-restored core, which can be important for large source lengths. Depending on the spatial structure of the source, this can lead to an order-one enhancement or suppression of baryon production.

Finite-mass effects also enter through the mass-gradient terms on the {\it lhs} of Eq.~\eqref{eq:fluid}. These terms become relevant only when the source is sufficiently narrow that the gradients compete with the interaction rates. A parametric estimate for the onset of this regime follows from
\begin{equation}
    v_{\mathrm{w}}\gamma_{\mathrm{w}}\,Q_i\,(m^2)'\ \sim\ \Gamma,
\end{equation}
where $Q_i$ stands for one of $Q_1$, $Q_2$, or $\bar R$. Taking $Q_i\simeq (1/10)\,T^{-2}$ as a representative value, $(m^2)'\sim y_t^2 F^2 T^2/l_{\rm source}$, and $\Gamma\simeq (1/30)\,T$, we obtain that the gradients are important for 
\begin{equation}
    l_{\rm source}\ \lesssim\ \Gamma^{-1}\,y_t^2F^2\,v_{\mathrm{w}}\gamma_{\mathrm{w}}
    \ \sim\ \frac{3}{T}\,v_{\mathrm{w}}\gamma_{\mathrm{w}}\,F^2y_t^2.
\end{equation}
For $F^2 v_{\mathrm{w}}\gamma_{\mathrm{w}} y_t^2\sim 1$, this threshold lies close to the WKB applicability requirement $l_{\rm source}\gg 1/T$, which restricts the region of parameter space where mass-gradient effects can be sizeable.

\begin{figure}
\center
\includegraphics[width=0.4 \textwidth]{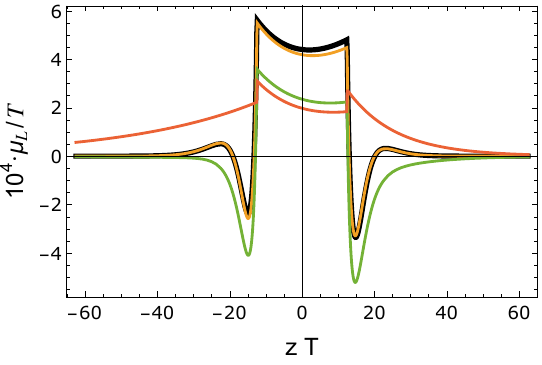}
\hspace{0.3cm}
\includegraphics[width=0.55 \textwidth]{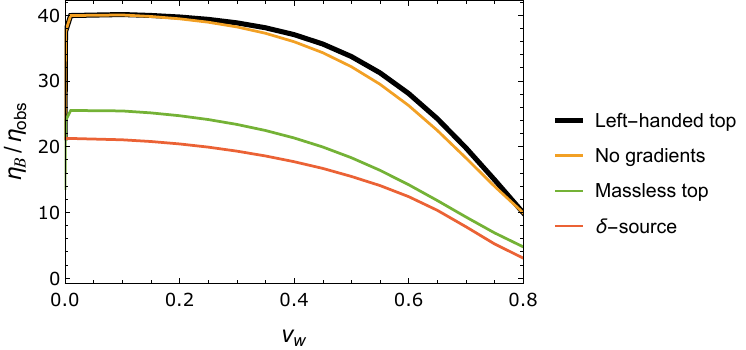}
\caption{Odd source in different approximation schemes, the thick black line includes all relevant effects. Left: $\mu_L$ profile for the benchmark model with $l_\text{wall} = 25/T$ and $l_\text{source} = 5/T$ and $v_w = 0.5$. Right: dependence of $\eta_B$ on the velocity of the wall.}
\label{fig:odd narrow source}
\end{figure}

\begin{figure}
\center
\includegraphics[width=0.4 \textwidth]{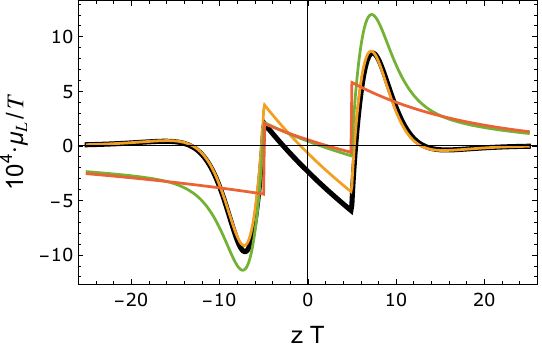}
\hspace{0.3cm}
\includegraphics[width=0.55 \textwidth]{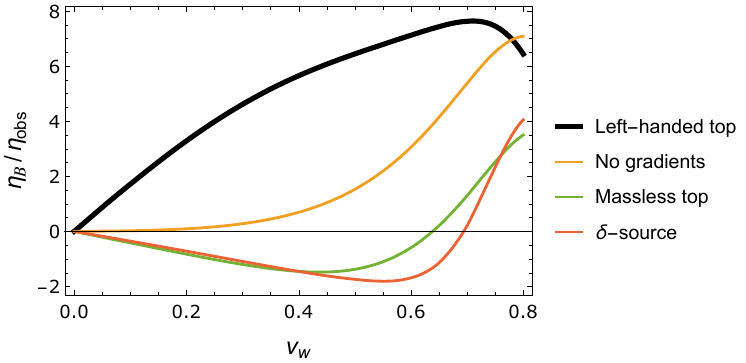}
\caption{Even source in different approximation schemes, the thick black line includes all relevant effects. Left: $\mu_L$ profile for the benchmark model with $l_\text{wall} = 10/T$ and $l_\text{source} = 4/T$ and $v_w = 0.5$. Right: dependence of $\eta_B$ on the velocity of the wall.}
\label{fig:even narrow source}
\end{figure}

\vspace{0.5 cm}

In Fig.~\ref{fig:odd narrow source} and~\ref{fig:even narrow source} we show examples of the $\mu_L$ and $\eta_B$ behaviour in the regimes without large hierarchy between $l_\Gamma$ and other relevant scales, for several approximate methods: 
\begin{itemize}
\item
complete solution with the left-handed top;
\item
solution neglecting mass gradient terms;
\item
setting the top mass to zero everywhere, except in the source;
\item
using a $\delta$-function-like source on top of the all above approximations, with the source normalized by
\begin{equation}\label{eq:sourcenorm}
\hat s_i = \int dz\, S_i(z),
\end{equation}
with the integral taken over the broken-phase region closest to $h=0$ and extending over a distance $l_\Gamma$.
\end{itemize}

As can be seen from the figures, the odd source yields fairly consistent results for both the chiral asymmetry and the baryon number across the various approximations and wall speeds. The discrepancy typically does not exceed a factor of two. It results from an overestimate of the contribution from the smaller (further from the core, see Fig.~\ref{fig:benchmark source}) peak of the source from massless top quark, and also from the delta-function approximation. The even source is more subtle: the baryon asymmetry is subject to the cancellations discussed in the previous section, and is therefore highly sensitive to order-one changes in the shape of the $\mu_L(z)$ profile. In this case, a fully numerical treatment becomes essential in the regimes without a sharp scale separation.
In particular, if, besides $l_{\rm source} \lesssim l_\Gamma$, also $l_{\rm wall} \lesssim l_\Gamma$ is realised, the simplified model predicts an additional suppression of $\eta_B$ even for moderate wall velocities. However, this is lifted when the full top mass dependence is included (cf. Fig.~\ref{fig:even narrow source}).

\begin{figure}
\center
\includegraphics[width=0.405 \textwidth]{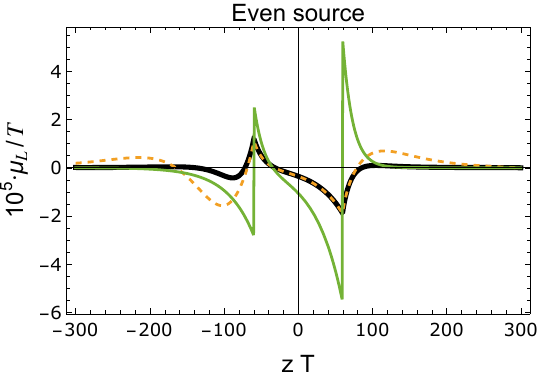}
\hspace{0.3cm}
\includegraphics[width=0.56 \textwidth]{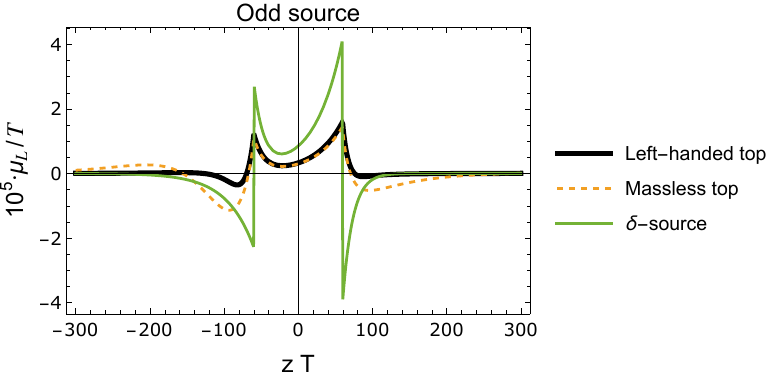}
\caption{Left (right) plot shows $\mu_L$ as generated by the benchmark model for an even (odd)  source with $l_\text{wall} = 120/T$, $l_\text{source} = 90/ T$, $v_w = 0.5$, $\Delta\theta = 1$ and $F=1.5$, where the thick black line again includes all relevant effects. See main text for more details on the approximations used.}
\label{fig:odd wide source}
\end{figure}

For comparison, Fig.~\ref{fig:odd wide source} shows the chiral asymmetry in the regime of large scale separation for several approximations. For both types of source, the $\delta$-function result is a factor of a few larger. This is explained by the fact that the normalization condition in Eq.~\eqref{eq:sourcenorm} overestimates the contribution from perturbations sourced at a distance $l_\Gamma$ from the EW symmetric region.

\vspace{0.5cm}
We can conclude that, even if the toy model does not capture some of the physical effects described in this section, it successfully predicts the baryon asymmetry generated by the full model, up to an order-one factor, in large part of the parameter space.

\begin{figure}
\center
\includegraphics[width=0.465\textwidth]
{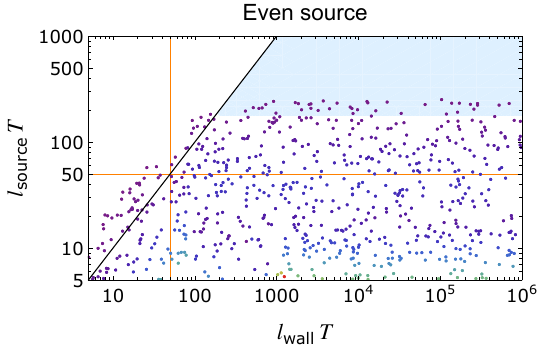}
\includegraphics[width=0.52\textwidth]{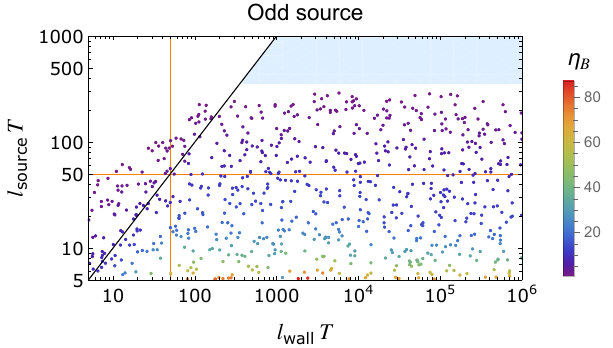}
\caption{Parameter space scan of the benchmark model for even and odd sources, see text for details.}
\label{fig:BenchmarkScan}
\end{figure}

\subsection{Parameter space scan}

We present the results of our numerical parameter-space scan in Fig.~\ref{fig:BenchmarkScan} for $F=1.5$, $v_{\mathrm{w}}=0.5$ and $\Delta \theta =1$, using the model with $t_L$ only. To evaluate the baryon asymmetry across the scan we employ a hybrid strategy, combining numerical solutions with analytic expressions in the regimes where they are reliable.

Colored points correspond to parameter choices that yield a sufficient baryon asymmetry; $\eta_B$ is shown in units of the observed value. The diagonal black line denotes $l_{\rm source}=l_{\rm wall}$. The vertical and horizontal orange lines indicate $l_{\rm wall}=l_\Gamma$ and $l_{\rm source}=l_\Gamma$, respectively, where we take $l_\Gamma\equiv 50/T$ as a representative perturbation decay length set by the dominant collision rates. Around these lines the analytic estimates can be inaccurate, as discussed above.

For $l_{\rm wall}T<80$ the computation is performed fully numerically. For $l_{\rm wall}T>80$, where the analytic treatment becomes more reliable, we instead use the analytic expression in Eq.~\eqref{eq:muint1}. In this regime, the normalization of the point-like sources is fixed numerically via Eq.~\eqref{eq:sourcenorm}.

Finally, the blue-shaded region indicates parameter space where the simple scaling estimate of Eq.~\eqref{eta:extsource} predicts an insufficient baryon asymmetry, within its regime of validity $l_{\rm wall}>l_{\rm source}$.

\subsection{Temperature dependence}\label{s:tedep}

We conclude the analysis of the generic domain-wall model by studying the temperature dependence of baryon production. To isolate effects associated with changing $T$ alone, it is convenient to keep all relevant length scales---$l_{\rm source}$, $l_{\rm wall}$, and $l_\Gamma$---fixed in units of the temperature. Varying $T$ then primarily amounts to varying the transition strength
$ F = {h(\pm\infty)}/{T}$.
The parameter $F$ affects the baryon asymmetry in three ways: it changes the profile $m_t(z)/T$, it modifies the weak sphaleron rate $\Gamma_{\rm sph}$, and it enters the washout term in Eq.~\eqref{eq:etagen}.

We first focus on the dependence through $m_t(z)/T$, which provides the dominant $F$ sensitivity. Increasing the top mass has three main effects. First, it enhances the CP-violating source. Since $S(z)\propto m^2$, the source amplitude near the electroweak-symmetric region grows parametrically as $F^2$. At the same time, the thermal functions $Q^{8}$ and $Q^{9}$ that enter the source become Boltzmann-suppressed at large $m/T$, shrinking the region where the source is efficient. Combining these effects, we find that for the top mass given by Eq.~\eqref{eq:benchmark t}, the integral of the source scales as  
\bea\label{eq:F scaling}
\int dz\,S(z)\ \propto
\begin{cases}
F^2 & F\sim 1,\\
const. & F\gg1.
\end{cases}
\eea

A second $F$-dependent effect is the propagation of quark perturbations in the plasma. Technically, increasing $F$ increases the eigenvalues $\lambda_\pm$ away from the wall, thereby shortening the diffusion length. This suppresses the contribution to $\mu_L$ sourced in regions that are not immediately adjacent to the electroweak-restored core. Because the source in those regions is already strongly Boltzmann suppressed at large $F$, we do not expect this effect to be quantitatively important.

Finally, larger $F$ enhances the relative importance of the mass-gradient terms discussed in Section~\ref{ss:physical effects beyond toy}. Much like the source, these terms scale as $F^2$ but are also Boltzmann suppressed by $\exp[-m_t(z)/T]$. Their net impact is typically an order-one modification of $\mu_L(z)$ inside the electroweak-restored region, together with the appearance of an oscillatory component in front of the leading edge of the wall. Numerically, this behaviour can lead to an apparent instability once $F$ becomes sufficiently large.

\begin{figure}
\center
\includegraphics[width=0.67 \textwidth]{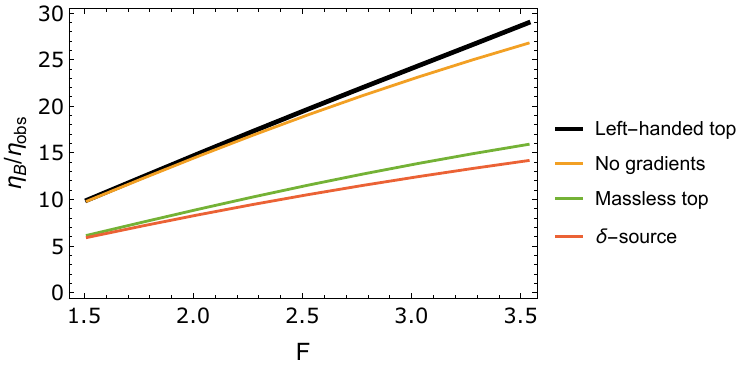}

\caption{Baryon asymmetry as function of $F$ for odd source with $l_\text{wall} = 30/T$, $l_\text{source} = 20/T$, $\Delta\theta = 1$ and $v_w =0.5$. 
}
\label{fig:benchmarkRegimeC}
\end{figure}

As expected, provided that the gradient terms remain subdominant, the $F$ dependence of $\eta_B$ is entirely driven by the $F$ dependence of the CP-violating source. This is illustrated in Fig.~\ref{fig:benchmarkRegimeC}, where we show the growth of $\eta_B$ with $F$ (for an odd source) while keeping all relevant length scales fixed in units of $T$.

The parameter $F$ also enters the electroweak sphaleron rate by controlling the degree of sphaleron suppression in the broken phase. Here, the only numerically relevant effect turns out to be the usual baryon asymmetry wash-out that requires $F\gtrsim 1$.

Finally, the washout term in Eq.~\eqref{eq:etagen} can, in principle, reduce the baryon asymmetry inside very thick walls. In practice, washout becomes important only when the corresponding exponent is of order unity, i.e.
\begin{equation}
\frac{l_{\rm wall}}{v_{\rm w}\gamma_{\rm w}}\,\Gamma_{\rm ws}\ \approx\ 1.
\end{equation}
Taking $\Gamma_{\rm ws}\approx 10^{-6}T$, this would require
\[
\frac{l_{\rm wall}T}{v_{\rm w}}\ \gtrsim\ 10^{6}.
\]
Such large values are never realized in the region of parameter space where baryon production is efficient. First, domain-wall--assisted EWBG is operative for relativistic wall velocities. Second, for $l_{\rm source}$ being an $\mathcal{O}(1)$ fraction of $l_{\rm wall}$, pushing $l_{\rm wall}$ to such values would strongly suppress gradients in the source and hence suppress $\eta_B$ according to Eq.~\eqref{eta:extsource}.

\section{Singlet-Extended SM}
\label{sec:singlet}

We now apply the considerations developed above to domain-wall-assisted electroweak baryogenesis in the singlet-extended Standard Model proposed in Ref.~\cite{Azzola:2024pzq}. In this setup, spontaneous breaking of the singlet $Z_2$ symmetry leads to the formation of domain walls. A negative Higgs--singlet cross-quartic coupling, $|H|^2S^2$, reduces the Higgs VEV inside the wall. If the variation of $S$ across the wall is sufficiently large, the wall core can enter a phase with restored electroweak symmetry. This provides the conditions for baryon number generation by a sweeping domain wall, which is the focus of the present work. Moreover, it was found in Ref.~\cite{Azzola:2024pzq} that just as in the single-field domain-wall case, the characteristic length scales of the fields' variation in this model are set by the inverse singlet mass, see Eq.~\eqref{eq:sprof}. A sufficiently light singlet therefore gives a wall core large enough to support unsuppressed weak sphaleron transitions, while also ensuring that the source varies slowly on the scale of the relevant particle de Broglie wavelengths. This justifies the use of the  transport formalism adopted in this paper.

Let us now specify the  relevant details of the model. The scalar potential is given by
\be
V = \mu_H^2 |H|^2 + \frac 1 2 \mu_S^2 S^2 + \lambda_H |H|^4 - \frac 1 2 |\lambda_{HS}| |H|^2 S^2 +  \frac 1 4 \lambda_S S^4.
\ee
The domain-wall configuration can be well approximated by the usual $\tanh$-profile for the scalar field, and the Higgs field tracking the local $S$-dependent minimum of its potential across the wall:
\bea
S(z) &=& v_S \tanh\left(\frac{m_S}{2} z\right), \label{eq:sprof}\\
h(z)^2 
&=& \left\{-\mu_H^2(T) + \frac{|\lambda_{HS}|}{2} S^2 \right\} / \lambda_H \\
&=& v_{h}^2 \left\{ 1 + \frac{\lambda_{HS} v_S^2}{2 \lambda_H v_{h}^2}  \left( S^2/v_S^2-1 \right) \right\} \\
&\equiv& (FT)^2 \left\{ 1 + \kappa \left(S^2/v_S^2-1 \right) \right\}, \label{eq:ssmh}
\eea
where the Higgs VEV $h^2$ is only non-zero when the {\it rhs} of Eq.~\eqref{eq:ssmh} is positive; $v_S$ and $m_S$ are singlet's VEV and physical mass in the minima of the potential; $v_h$ is a temperature-dependent VEV of the Higgs field in the broken phase, and $F$ is the phase transition strength. In the last equation we introduced parameter $\kappa$ which defines how quickly the Higgs VEV reacts to the $S$ variation, and is related to the fine-tuning of the Higgs potential.

\begin{figure}
\center
\includegraphics[width=0.67 \textwidth]{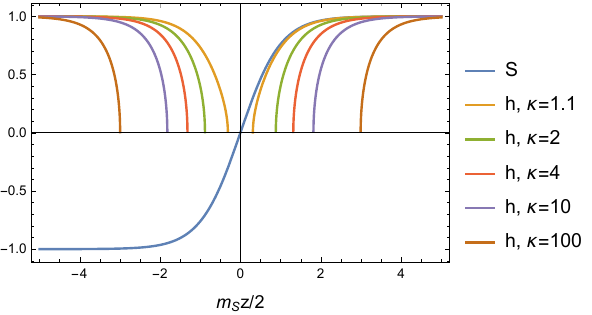}
\caption{Dependence of the field profiles on the Higgs VEV tuning $\kappa$.}
\label{fig:tuning}
\end{figure}

\vspace{0.5cm}
{\bf Wall width.} We will define the wall width $l_{\text{wall}}$ as the length of the region with vanishing Higgs VEV. Using Eq.~\eqref{eq:ssmh} we find that the corresponding condition $h(z=l_{\text{wall}}/2)=0$ translates into
\be\label{eq:lwallssm}
l_{\text{wall}} 
= \frac{4}{m_S} \text{arctanh} \sqrt{1 - 1/\kappa} 
\simeq 
\frac{4}{m_S} \left(\log 2 + \log(\kappa)/2\right)
\ee
where in the last step we show large-$\kappa$ expansion. This means that the wall core can be enhanced with respect to the naive $\sim 1/m_S$ value if the Higgs mass is significantly tuned. In Fig.~\ref{fig:tuning} we show the comparison of the normalized Higgs and $S$ profiles for different $\kappa$.

\vspace{0.5cm}
{\bf CPV source.}
Following Ref.~\cite{Azzola:2024pzq}, we use a complex top quark mass as a source of CP-violation. It originates from the interaction 
\be \label{eq: CPV}
    \mathcal{L} \supset - \frac{y_t}{\sqrt{2}} h \bar{t}_L t_R \left( 1 - i \left(\frac{S}{f} \right)^{n_\text{CP}} \right) + h.c.,
\ee
where $n_{CP} = 1(2)$ results in odd(even) CPV-source, and $f$ is a new scale associated with the singlet, taken to be of the order of $v_S$. In particular, from Eq.~\eqref{eq: CPV}, it follows that 
\begin{equation}
    \theta_t = \arg(m_t(z)) = - \arctan \left[ \left( \frac{S(z)}{f} \right)^{n_{CP}} \right],
\end{equation}
from which one can see that the ratio $v_S/f$ controls the total change in the top quark mass phase, and is therefore analogous to $\Delta\theta$ of the previous section.

\vspace{0.5cm}
{\bf Source width.}
The source~\eqref{eq: CPV} depends on $h$ and $S$ profiles, which are essentially functions of $\tanh[m_S z/2]$, hence, independently of the tuning, the typical scale of the source variation is  
\be
l_{\text{source}} \simeq 2/m_S.
\ee
On top of this, the source experiences a Boltzmann suppression in regions where the top quark mass is much larger than temperature, regardless of mass and phase gradients. In this region the CPV force on top quarks can be sizeable, but almost no top quarks are present to experience such force. This effect is encompassed by the $Q_8,Q_9$ coefficients in Eq.~\eqref{eq:cpvfull}. As a result, for low temperatures, the width of the non-vanishing source region scales as $\sim1/(m_SF^2)$.

\vspace{0.5cm}
{\bf Source magnitude.}
The leading-order CP-violating sources entering the transport equations are proportional to either $(m_t^2)'\theta_t'$ or $m_t^2 \theta_t''$, see Eq.~\eqref{eq:cpvfull}. It can be readily verified that at $z = \pm l_{\text{wall}}/2$ where a non-vanishing $m_t$ appears, and the variation is maximized, the following parametric estimates hold:
\bea
(m_t^2)'/m_t^2 &\propto& m_S, \\
\theta_t' &\propto& m_S/\kappa.
\eea
Therefore only a mild tuning of the Higgs mass $\kappa $ can be allowed before the CP-violation gets unacceptably suppressed. This, in turn, means that the wall width defined in Eq.~\eqref{eq:lwallssm} cannot be more than a factor of a few larger than the naive estimate $4/m_S$. To avoid this conclusion one would need to use a different type of CPV source, either having a different functional form, or relying on other fields than $S$ to generate the varying phase. In summary, the source length and the wall length turn out to be rather rigidly related in the minimal set-up that we analyze.

\begin{figure}
\center
\includegraphics[width=0.47 \textwidth]{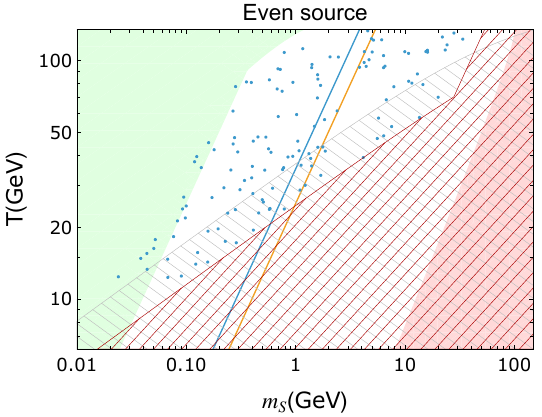}
\includegraphics[width=0.47 \textwidth]{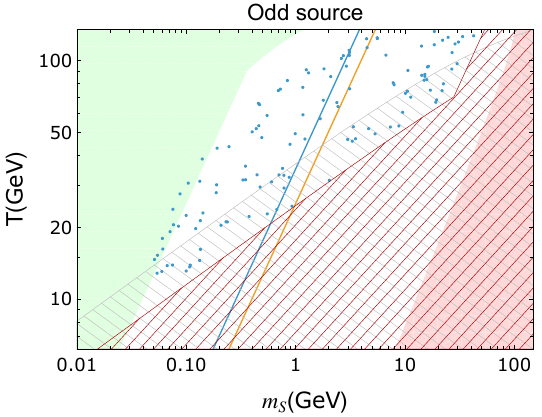}
\caption{
Scatter plots of parameter points yielding sufficient baryon asymmetry for a single wall passage in the singlet-extended SM. The left (right) panel corresponds to the even (odd) source. Green and red shaded regions are expected to be devoid of viable points according to our analytic estimates: in the green region large wall widths suppress gradients and hence CPV, and in the red region the asymmetry is suppressed by Higgs tuning required by the condition \( l_{\text{wall}} > l_{\text{sphaleron}} \). The gray hatched area indicates where the asymmetry is computed analytically because of the numerical instabilities due to mass-gradient terms. The red hatched region marks where gradient expansion, Eq.~\eqref{eq: WKB condition}, or approximate kinetic equilibrium,  Eq.~\eqref{eq: quasi equilibrium breaks}, fails. The blue (orange) diagonal line corresponds to \( l_{\text{wall}} T = 80 \) (\( l_{\text{source}} = l_{\Gamma} \)).}
\label{fig: SSMB}
\end{figure}

\vspace{0.5cm}
{\bf Baryon asymmetry.}
We now present the results of our computation of the baryon asymmetry in the singlet-extended Standard Model of Ref.~\cite{Azzola:2024pzq}. We use a single-top quark model discussed in the previous section. The results are shown in Fig.~\ref{fig: SSMB} as a function of the singlet mass and the temperature at which the domain-wall evolution occurs; these parameters are randomly scanned. Blue points correspond to parameter choices that yield a baryon-to-entropy ratio $\eta_B \geq 10^{-10}$.

Throughout this analysis we fix the wall velocity to $v_{\mathrm{w}}=0.5$ and the CP-violating phase variation to $\Delta\theta=1$. 

As for the benchmark model scan in Fig.~\ref{fig:BenchmarkScan}, in the region $l_{\rm wall}T<80$ (to the right of the blue line in Fig.~\ref{fig: SSMB}), we compute the baryon asymmetry fully numerically; for $l_{\rm wall}T>80$ we instead use the analytic expression in Eq.~\eqref{eq:muint1}. 
In addition, the orange line in Fig.~\ref{fig: SSMB} corresponds to  $l_\text{source}=l_\Gamma$, which serves to connect the scanned parameter space with the parametric regimes discussed in earlier sections.

The strength of the phase transition is fixed using an approximate SM-like temperature dependence of the Higgs VEV,
\begin{equation} \label{eq: F SM}
F = \frac{h}{T}, \qquad 
h = v_{h,\mathrm{SM}}\sqrt{1-(T/T_c)^2}, \qquad 
T_c = 160~\mathrm{GeV}.
\end{equation}

The parameter $\kappa$, which controls the wall thickness, is chosen as
\[
\kappa=\max\!\left[2,\kappa_{\rm min}\right],
\]
where $\kappa_{\rm min}$ denotes the minimal tuning required to ensure $l_{\rm wall}>l_{\rm sphaleron}$ (see Ref.~\cite{Azzola:2024pzq}). The sphaleron length is given by
\[
l_{\rm sphaleron}=\frac{4\pi}{g^2}\frac{1}{T}.
\]
At the same time, we impose $\kappa\geq 2$ in order to guarantee a sizeable region of electroweak symmetry restoration inside the wall, as illustrated in Fig.~\ref{fig:tuning}.

The color-shaded regions in Fig.~\ref{fig: SSMB} indicate parameter regions excluded on the basis of analytic estimates. The green region corresponds to $\eta_B<10^{-10}$ according to Eq.~\eqref{eta:extsource}, which is valid for $l_\Gamma\ll l_{\rm source},\,l_{\rm wall}$. This estimate is modified by the inclusion of $F(T)$ dependence, as motivated in Section~\ref{s:tedep}, and we fix $l_\Gamma=50/T$. The red region corresponds to $\eta_B<10^{-10}$ as predicted by Eq.~\eqref{eta:thinwall}, valid for $l_\Gamma\gg l_{\rm source},\,l_{\rm wall}$, again including the factor $F(T)$ as well as a $1/\kappa$ suppression factor arising from the reduced source strength.

The gray hatched region denotes the parameter space where gradient terms in the transport equations become important, satisfying
\begin{equation}
v_{\rm w}\gamma_{\rm w}\,Q_i\,(m^2)' \gtrsim \Gamma,
\end{equation}
which approximately corresponds to the condition  $F \gtrsim 0.76\sqrt{l_\text{source} T}$. In this region, $\eta_B$ is evaluated only using the analytic expressions, as the numerical solution exhibits instabilities.

Furthermore, the fluid ansatz in Eq.\eqref{eq:fluan} assumes that the top quarks are sufficiently close to kinetic equilibrium. We require this equilibration to occur within the portion of the wall where the source predicted by the ansatz is active and not Boltzmann suppressed, which we estimate by the condition $m_t \lesssim 2T$. Following Ref.~\cite{Joyce:1994zt}, we take the kinetic equilibration length of the top quarks to be approximately $15 v_{\rm w}/T$. The width of the transition region near either edge of the wall over which the top mass changes between zero and $2T$ is estimated as
\be
\Delta z_{m_t<2T} \sim 2l_{\rm source} \left[\frac{2T}{m_t({\rm bulk})}\right]^2,
\qquad
l_{\rm source}\sim {2}/{m_S}.
\ee
Requiring the equilibration length to be smaller than this active region gives
\be \label{eq: quasi equilibrium breaks}
m_S \lesssim \frac{2}{v_{\rm w}}\,
\frac{T^3}{h({\rm bulk})^2} = \frac{2}{v_{\rm w}}\,
\frac{T}{F(T)^2}.
\ee

A second constraint on the validity of our treatment arises from the WKB expansion. The semiclassical approximation requires the mass profile to vary on length scales larger than the thermal de Broglie wavelength. We therefore impose
\begin{equation}
\label{eq: WKB condition}
l_{\rm source} \sim \frac{2}{m_S} < \frac{5}{T}.
\end{equation}
The red hatched region in Fig.~\ref{fig: SSMB} denotes the parameter space where at least one of the validity conditions, Eqs.~\eqref{eq: quasi equilibrium breaks} and \eqref{eq: WKB condition}, is violated.

\begin{figure}
\center
\includegraphics[width=0.8 \textwidth]{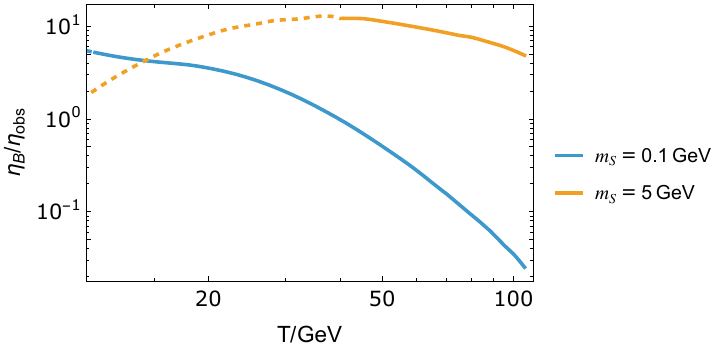}

\caption{Temperature dependence of the baryon asymmetry in the singlet-extended SM, for $v_w = 0.5$, $\Delta\theta =1$, $\kappa = 2$, $n_\text{CP} = 2$ and $F(T)$ from Eq.~\eqref{eq: F SM}. The suppression from the sphaleron size is modeled as a multiplicative factor $\exp(-l_\text{sphaleron}(T)/l_\text{wall})$. Results are computed with the $\delta$-source model. Dashed lines correspond to temperatures where the condition on top quark thermalisation length \eqref{eq: quasi equilibrium breaks} is not met.}
\label{fig: full T}
\end{figure}

\paragraph{Temperature dependence} 

We now discuss the temperature dependence of the baryon asymmetry generated by a single wall passage in the singlet-extended SM. The mechanism considered here requires electroweak symmetry to be broken in the bulk of the Universe, while the singlet domain walls have already formed and retain an electroweak-symmetric core. It can therefore operate only after both the electroweak and $Z_2$-breaking phase transitions, and before the domain walls decay. Consequently, the upper end of the relevant temperature interval is determined by the later of two conditions: the onset of sufficiently strong electroweak breaking, $h/T \gtrsim 1$, and the model-dependent temperature at which the $Z_2$ symmetry is broken, which depends on the details of the UV completion~\cite{Azzola:2024pzq}. The lower end of the interval is set by the size of the explicit $Z_2$ breaking that eventually drives the decay of the domain-wall network. The walls must decay before they dominate the energy density of the Universe~\cite{Azzola:2024pzq}, and furthermore the relevant baryon-asymmetry production must occur before BBN.

Within this interval, the baryon-asymmetry production rate generically increases as the temperature decreases. The precise scaling is controlled by the transport dynamics discussed in the previous sections, as well as by the wall width, which is set by the singlet mass. This growth eventually ceases when the wall width becomes smaller than the sphaleron size, leading to a suppression of the generated asymmetry. As a representative example, Fig.~\ref{fig: full T} shows the full temperature dependence of the baryon asymmetry for two singlet masses near the opposite edges of the allowed range, for temperatures between $5$ and $110~{\rm GeV}$.

For the lighter singlet mass, the asymmetry scales as $\eta_B \sim T^{-4}$ at high temperatures. This behavior follows from the wide-wall/source scaling summarized in Table~\ref{tab:summary regimes}, supplemented by the enhancement of the CP-violating force proportional to $(m_t/T)^2 \propto F^2 \propto T^{-2}$. At later times, once $F$ becomes large, the Boltzmann suppression of the top-quark density near the electroweak-symmetric core becomes important. This weakens the $(l_\Gamma/l_{\rm source})^2$ scaling and leads instead to an approximate $\eta_B \sim T^{-1}$ behavior at low temperatures. For larger singlet masses, the high-temperature scaling is fully driven by the $F^2$ enhancement of the CP-violating force.

It is worth emphasizing that our results show that efficient baryon-asymmetry production in our scenario can occur at temperatures as low as $\sim 10$~GeV. This is significantly below the typical scale of standard electroweak baryogenesis in the singlet-extended SM, where the relevant temperature is usually of order 100 GeV~\cite{Beniwal:2017eik}.

\section{Conclusions} \label{s:conc}

We have investigated electroweak baryogenesis from domain walls with electroweak-symmetric cores in the thick-wall regime. In this setup, CP-violating semiclassical forces generate particle asymmetries that diffuse through the plasma and bias weak sphalerons. Compared with conventional bubble-wall electroweak baryogenesis, a distinctive feature of the domain-wall case is that the wall contains two interfaces, whose contributions to the final baryon asymmetry can interfere.

Our main result is that the efficiency of baryon production is governed by the hierarchy between the wall width, the width of the CP-violating source, and the interaction or diffusion length. This hierarchy controls both the transport dynamics and the degree of cancellation between the two faces of the wall. In particular, we find that sources that are even under reversal of the wall orientation are generically more strongly suppressed, while odd sources can lead to a parametrically larger asymmetry.

A simplified one-species description captures these effects in a transparent way and reproduces the full transport analysis over a broad region of parameter space. The comparison with the full network moreover identifies the regimes in which the reduced description is reliable and clarifies the temperature dependence of the generated asymmetry.

Applying this framework to the singlet-extended Standard Model, we find that successful baryogenesis favors a light singlet with mass in the approximate range
\[
m_S \sim {\cal O}(10)\,\mathrm{MeV} - {\cal O}(10)\,\mathrm{GeV},
\]
together with a small but non-zero singlet--Higgs mixing. Compared with the conclusions of~\cite{Azzola:2024pzq}, our more detailed analysis disfavors the very light singlet regime, in which the wall becomes too wide for the CP-violating source to generate a sufficiently large asymmetry. This scenario is predictive and experimentally testable through a combination of collider and fixed-target searches~\cite{CHARM:1985anb,Banerjee:2020kww,Clarke:2013aya,Fuchs:2020cmm,Carena:2022yvx,ATLAS:2019nkf}, EDM measurements~\cite{ACME:2018yjb}, and cosmological~\cite{Fradette:2018hhl,DEramo:2024lsk} and astrophysical~\cite{Turner:1987by,Burrows:1988ah,Hardy:2024gwy,Caputo:2022mah,Diamond:2023cto} probes. 


Taken together, these results strengthen the case for domain-wall--assisted electroweak baryogenesis and show that its viability is controlled by a small set of physically transparent ingredients: transport, scale hierarchies, and interference between the two faces of the wall.

\vspace{1.cm}
{\bf Acknowledgements}

We would like to thank Stefan Stelzl, Aleksandr Azatov,  Bruno Missoni and Giulio Barni for useful discussions.   
This work was partially supported by the Collaborative Research Center SFB1258, the Munich Institute for Astro-, Particle and BioPhysics (MIAPbP), the Excellence Cluster ORIGINS, which is funded by the Deutsche Forschungsgemeinschaft (DFG, German Research Foundation) under Germany’s Excellence Strategy – EXC-2094-390783311.

%
%

\bibliographystyle{JHEP} 
\bibliography{biblio} 

\providecommand{\href}[2]{#2}\begingroup\raggedright\begin{thebibliography}{10}

\bibitem{Shaposhnikov:1987tw}
M.~E. Shaposhnikov, \emph{{Baryon Asymmetry of the Universe in Standard
  Electroweak Theory}},
  \href{https://doi.org/10.1016/0550-3213(87)90127-1}{\emph{Nucl. Phys. B}
  {\bfseries 287} (1987) 757--775}.

\bibitem{Cohen:1990it}
A.~G. Cohen, D.~B. Kaplan and A.~E. Nelson, \emph{{Baryogenesis at the weak
  phase transition}},
  \href{https://doi.org/10.1016/0550-3213(91)90395-E}{\emph{Nucl. Phys. B}
  {\bfseries 349} (1991) 727--742}.

\bibitem{Morrissey:2012db}
D.~E. Morrissey and M.~J. Ramsey-Musolf, \emph{{Electroweak baryogenesis}},
  \href{https://doi.org/10.1088/1367-2630/14/12/125003}{\emph{New J. Phys.}
  {\bfseries 14} (2012) 125003},
  [\href{https://arxiv.org/abs/1206.2942}{{\ttfamily 1206.2942}}].

\bibitem{Cline:2006ts}
J.~M. Cline, \emph{{Baryogenesis}},  in \emph{{Les Houches Summer School -
  Session 86: Particle Physics and Cosmology: The Fabric of Spacetime}}, 9,
  2006, \href{https://arxiv.org/abs/hep-ph/0609145}{{\ttfamily
  hep-ph/0609145}}.

\bibitem{Sakharov:1967dj}
A.~D. Sakharov, \emph{{Violation of CP Invariance, C asymmetry, and baryon
  asymmetry of the universe}},
  \href{https://doi.org/10.1070/PU1991v034n05ABEH002497}{\emph{Pisma Zh. Eksp.
  Teor. Fiz.} {\bfseries 5} (1967) 32--35}.

\bibitem{Brandenberger:1994mq}
R.~H. Brandenberger, A.-C. Davis, T.~Prokopec and M.~Trodden, \emph{{Local and
  nonlocal defect mediated electroweak baryogenesis}},
  \href{https://doi.org/10.1103/PhysRevD.53.4257}{\emph{Phys. Rev. D}
  {\bfseries 53} (1996) 4257--4266},
  [\href{https://arxiv.org/abs/hep-ph/9409281}{{\ttfamily hep-ph/9409281}}].

\bibitem{Abel:1995uc}
S.~A. Abel and P.~L. White, \emph{{Baryogenesis from domain walls in the
  next-to-minimal supersymmetric standard model}},
  \href{https://doi.org/10.1103/PhysRevD.52.4371}{\emph{Phys. Rev. D}
  {\bfseries 52} (1995) 4371--4379},
  [\href{https://arxiv.org/abs/hep-ph/9505241}{{\ttfamily hep-ph/9505241}}].

\bibitem{Brandenberger:2005bx}
R.~H. Brandenberger, W.~Kelly and M.~Yamaguchi, \emph{{Electroweak baryogenesis
  with embedded domain walls}},
  \href{https://doi.org/10.1143/PTP.117.823}{\emph{Prog. Theor. Phys.}
  {\bfseries 117} (2007) 823--834},
  [\href{https://arxiv.org/abs/hep-ph/0503211}{{\ttfamily hep-ph/0503211}}].

\bibitem{Bai:2021xyf}
Y.~Bai, J.~Berger, M.~Korwar and N.~Orlofsky, \emph{{Catalyzed baryogenesis}},
  \href{https://doi.org/10.1007/JHEP10(2021)147}{\emph{JHEP} {\bfseries 10}
  (2021) 147}, [\href{https://arxiv.org/abs/2106.12589}{{\ttfamily
  2106.12589}}].

\bibitem{Sassi:2024cyb}
M.~Y. Sassi and G.~Moortgat-Pick, \emph{{Electroweak Symmetry Restoration in
  the N2HDM via Domain Walls}},
  \href{https://arxiv.org/abs/2407.14468}{{\ttfamily 2407.14468}}.

\bibitem{Schroder:2024gsi}
T.~Schr\"oder and R.~Brandenberger, \emph{{Embedded domain walls and
  electroweak baryogenesis}},
  \href{https://doi.org/10.1103/PhysRevD.110.043516}{\emph{Phys. Rev. D}
  {\bfseries 110} (2024) 043516},
  [\href{https://arxiv.org/abs/2404.13035}{{\ttfamily 2404.13035}}].

\bibitem{Klipfel:2026nzx}
A.~P. Klipfel, M.~Vanvlasselaer, S.~Trifinopoulos and D.~I. Kaiser,
  \emph{{Baryogenesis from Exploding Primordial Black Holes}},
  \href{https://arxiv.org/abs/2603.29024}{{\ttfamily 2603.29024}}.

\bibitem{Azzola:2024pzq}
J.~Azzola, O.~Matsedonskyi and A.~Weiler, \emph{{Minimal electroweak
  baryogenesis via domain walls}},
  \href{https://doi.org/10.1007/JHEP04(2025)103}{\emph{JHEP} {\bfseries 04}
  (2025) 103}, [\href{https://arxiv.org/abs/2412.10495}{{\ttfamily
  2412.10495}}].

\bibitem{Press:1989yh}
W.~H. Press, B.~S. Ryden and D.~N. Spergel, \emph{{Dynamical Evolution of
  Domain Walls in an Expanding Universe}},
  \href{https://doi.org/10.1086/168151}{\emph{Astrophys. J.} {\bfseries 347}
  (1989) 590--604}.

\bibitem{Espinosa:2011ax}
J.~R. Espinosa, T.~Konstandin and F.~Riva, \emph{{Strong Electroweak Phase
  Transitions in the Standard Model with a Singlet}},
  \href{https://doi.org/10.1016/j.nuclphysb.2011.09.010}{\emph{Nucl. Phys. B}
  {\bfseries 854} (2012) 592--630},
  [\href{https://arxiv.org/abs/1107.5441}{{\ttfamily 1107.5441}}].

\bibitem{Espinosa:2011eu}
J.~R. Espinosa, B.~Gripaios, T.~Konstandin and F.~Riva, \emph{{Electroweak
  Baryogenesis in Non-minimal Composite Higgs Models}},
  \href{https://doi.org/10.1088/1475-7516/2012/01/012}{\emph{JCAP} {\bfseries
  01} (2012) 012}, [\href{https://arxiv.org/abs/1110.2876}{{\ttfamily
  1110.2876}}].

\bibitem{Ellis:2022lft}
J.~Ellis, M.~Lewicki, M.~Merchand, J.~M. No and M.~Zych, \emph{{The scalar
  singlet extension of the Standard Model: gravitational waves versus
  baryogenesis}}, \href{https://doi.org/10.1007/JHEP01(2023)093}{\emph{JHEP}
  {\bfseries 01} (2023) 093},
  [\href{https://arxiv.org/abs/2210.16305}{{\ttfamily 2210.16305}}].

\bibitem{Carena:2019une}
M.~Carena, Z.~Liu and Y.~Wang, \emph{{Electroweak phase transition with
  spontaneous Z$_{2}$-breaking}},
  \href{https://doi.org/10.1007/JHEP08(2020)107}{\emph{JHEP} {\bfseries 08}
  (2020) 107}, [\href{https://arxiv.org/abs/1911.10206}{{\ttfamily
  1911.10206}}].

\bibitem{Beniwal:2017eik}
A.~Beniwal, M.~Lewicki, J.~D. Wells, M.~White and A.~G. Williams,
  \emph{{Gravitational wave, collider and dark matter signals from a scalar
  singlet electroweak baryogenesis}},
  \href{https://doi.org/10.1007/JHEP08(2017)108}{\emph{JHEP} {\bfseries 08}
  (2017) 108}, [\href{https://arxiv.org/abs/1702.06124}{{\ttfamily
  1702.06124}}].

\bibitem{ACME:2018yjb}
{\scshape ACME} collaboration, V.~Andreev et~al., \emph{{Improved limit on the
  electric dipole moment of the electron}},
  \href{https://doi.org/10.1038/s41586-018-0599-8}{\emph{Nature} {\bfseries
  562} (2018) 355--360}.

\bibitem{Gouttenoire:2025ofv}
Y.~Gouttenoire, S.~F. King, R.~Roshan, X.~Wang, G.~White and M.~Yamazaki,
  \emph{{Cosmological consequences of domain walls biased by quantum gravity}},
  \href{https://doi.org/10.1103/7zmx-v16z}{\emph{Phys. Rev. D} {\bfseries 112}
  (2025) 075007}, [\href{https://arxiv.org/abs/2501.16414}{{\ttfamily
  2501.16414}}].

\bibitem{DEramo:2024lsk}
F.~D'Eramo, A.~Tesi and V.~Vaskonen, \emph{{Irreducible cosmological
  backgrounds of a real scalar with a broken symmetry}},
  \href{https://arxiv.org/abs/2407.19997}{{\ttfamily 2407.19997}}.

\bibitem{Blasi:2025tmn}
S.~Blasi, A.~Mariotti, A.~Rase and M.~Vanvlasselaer, \emph{{Domain walls in the
  scaling regime: Equal Time Correlator and Gravitational Waves}},
  \href{https://arxiv.org/abs/2511.16649}{{\ttfamily 2511.16649}}.

\bibitem{Blasi:2023rqi}
S.~Blasi, R.~Jinno, T.~Konstandin, H.~Rubira and I.~Stomberg,
  \emph{{Gravitational waves from defect-driven phase transitions: domain
  walls}}, \href{https://doi.org/10.1088/1475-7516/2023/10/051}{\emph{JCAP}
  {\bfseries 10} (2023) 051},
  [\href{https://arxiv.org/abs/2302.06952}{{\ttfamily 2302.06952}}].

\bibitem{Azzola:2026mbq}
J.~Azzola, O.~Matsedonskyi and A.~Weiler, \emph{{Inverse Electroweak
  Baryogenesis}},  \href{https://arxiv.org/abs/2603.20414}{{\ttfamily
  2603.20414}}.

\bibitem{Bagherian:2025puf}
H.~Bagherian, M.~Ekhterachian and S.~Stelzl, \emph{{The Bearable Inhomogeneity
  of the Baryon Asymmetry}},
  \href{https://arxiv.org/abs/2505.15904}{{\ttfamily 2505.15904}}.

\bibitem{Azatov:2026sdm}
A.~Azatov and B.~Missoni, \emph{{Bounds from D/H on baryogenesis models}},
  \href{https://arxiv.org/abs/2604.11203}{{\ttfamily 2604.11203}}.

\bibitem{Avelino:2008ve}
P.~P. Avelino, C.~J. A.~P. Martins, J.~Menezes, R.~Menezes and J.~C. R.~E.
  Oliveira, \emph{{Dynamics of domain wall networks with junctions}},
  \href{https://doi.org/10.1103/PhysRevD.78.103508}{\emph{Phys. Rev. D}
  {\bfseries 78} (2008) 103508},
  [\href{https://arxiv.org/abs/0807.4442}{{\ttfamily 0807.4442}}].

\bibitem{Cline:2020jre}
J.~M. Cline and K.~Kainulainen, \emph{{Electroweak baryogenesis at high bubble
  wall velocities}},
  \href{https://doi.org/10.1103/PhysRevD.101.063525}{\emph{Phys. Rev. D}
  {\bfseries 101} (2020) 063525},
  [\href{https://arxiv.org/abs/2001.00568}{{\ttfamily 2001.00568}}].

\bibitem{Joyce:1994zt}
M.~Joyce, T.~Prokopec and N.~Turok, \emph{{Nonlocal electroweak baryogenesis.
  Part 2: The Classical regime}},
  \href{https://doi.org/10.1103/PhysRevD.53.2958}{\emph{Phys. Rev. D}
  {\bfseries 53} (1996) 2958--2980},
  [\href{https://arxiv.org/abs/hep-ph/9410282}{{\ttfamily hep-ph/9410282}}].

\bibitem{Cline:2000nw}
J.~M. Cline, M.~Joyce and K.~Kainulainen, \emph{{Supersymmetric electroweak
  baryogenesis}},
  \href{https://doi.org/10.1088/1126-6708/2000/07/018}{\emph{JHEP} {\bfseries
  07} (2000) 018}, [\href{https://arxiv.org/abs/hep-ph/0006119}{{\ttfamily
  hep-ph/0006119}}].

\bibitem{Fromme:2006wx}
L.~Fromme and S.~J. Huber, \emph{{Top transport in electroweak baryogenesis}},
  \href{https://doi.org/10.1088/1126-6708/2007/03/049}{\emph{JHEP} {\bfseries
  03} (2007) 049}, [\href{https://arxiv.org/abs/hep-ph/0604159}{{\ttfamily
  hep-ph/0604159}}].

\bibitem{Dorsch:2021ubz}
G.~C. Dorsch, S.~J. Huber and T.~Konstandin, \emph{{On the wall velocity
  dependence of electroweak baryogenesis}},
  \href{https://doi.org/10.1088/1475-7516/2021/08/020}{\emph{JCAP} {\bfseries
  08} (2021) 020}, [\href{https://arxiv.org/abs/2106.06547}{{\ttfamily
  2106.06547}}].

\bibitem{Bodeker:1999gx}
D.~Bodeker, G.~D. Moore and K.~Rummukainen, \emph{{Chern-Simons number
  diffusion and hard thermal loops on the lattice}},
  \href{https://doi.org/10.1103/PhysRevD.61.056003}{\emph{Phys. Rev. D}
  {\bfseries 61} (2000) 056003},
  [\href{https://arxiv.org/abs/hep-ph/9907545}{{\ttfamily hep-ph/9907545}}].

\bibitem{Li:2025kyo}
X.-X. Li, M.~J. Ramsey-Musolf, T.~V.~I. Tenkanen and Y.~Wu, \emph{{An Effective
  Sphaleron Awakens}},  \href{https://arxiv.org/abs/2506.01585}{{\ttfamily
  2506.01585}}.

\bibitem{Bruggisser:2017lhc}
S.~Bruggisser, T.~Konstandin and G.~Servant, \emph{{CP-violation for
  Electroweak Baryogenesis from Dynamical CKM Matrix}},
  \href{https://doi.org/10.1088/1475-7516/2017/11/034}{\emph{JCAP} {\bfseries
  11} (2017) 034}, [\href{https://arxiv.org/abs/1706.08534}{{\ttfamily
  1706.08534}}].

\bibitem{Kainulainen:2002th}
K.~Kainulainen, T.~Prokopec, M.~G. Schmidt and S.~Weinstock,
  \emph{{Semiclassical force for electroweak baryogenesis: Three-dimensional
  derivation}}, \href{https://doi.org/10.1103/PhysRevD.66.043502}{\emph{Phys.
  Rev. D} {\bfseries 66} (2002) 043502},
  [\href{https://arxiv.org/abs/hep-ph/0202177}{{\ttfamily hep-ph/0202177}}].

\bibitem{Cline:2021dkf}
J.~M. Cline and B.~Laurent, \emph{{Electroweak baryogenesis from light fermion
  sources: A critical study}},
  \href{https://doi.org/10.1103/PhysRevD.104.083507}{\emph{Phys. Rev. D}
  {\bfseries 104} (2021) 083507},
  [\href{https://arxiv.org/abs/2108.04249}{{\ttfamily 2108.04249}}].

\bibitem{Kainulainen:2024qpm}
K.~Kainulainen and N.~Venkatesan, \emph{{Systematic moment expansion for
  electroweak baryogenesis}},
  \href{https://doi.org/10.1088/1475-7516/2024/08/058}{\emph{JCAP} {\bfseries
  08} (2024) 058}, [\href{https://arxiv.org/abs/2407.13639}{{\ttfamily
  2407.13639}}].

\bibitem{Arnold:1987mh}
P.~B. Arnold and L.~D. McLerran, \emph{{Sphalerons, Small Fluctuations and
  Baryon Number Violation in Electroweak Theory}},
  \href{https://doi.org/10.1103/PhysRevD.36.581}{\emph{Phys. Rev. D} {\bfseries
  36} (1987) 581}.

\bibitem{CHARM:1985anb}
{\scshape CHARM} collaboration, F.~Bergsma et~al., \emph{{Search for Axion Like
  Particle Production in 400-{GeV} Proton - Copper Interactions}},
  \href{https://doi.org/10.1016/0370-2693(85)90400-9}{\emph{Phys. Lett. B}
  {\bfseries 157} (1985) 458--462}.

\bibitem{Banerjee:2020kww}
A.~Banerjee, H.~Kim, O.~Matsedonskyi, G.~Perez and M.~S. Safronova,
  \emph{{Probing the Relaxed Relaxion at the Luminosity and Precision
  Frontiers}}, \href{https://doi.org/10.1007/JHEP07(2020)153}{\emph{JHEP}
  {\bfseries 07} (2020) 153},
  [\href{https://arxiv.org/abs/2004.02899}{{\ttfamily 2004.02899}}].

\bibitem{Clarke:2013aya}
J.~D. Clarke, R.~Foot and R.~R. Volkas, \emph{{Phenomenology of a very light
  scalar (100 MeV \ensuremath{<} $m_h$ \ensuremath{<} 10 GeV) mixing with the
  SM Higgs}}, \href{https://doi.org/10.1007/JHEP02(2014)123}{\emph{JHEP}
  {\bfseries 02} (2014) 123},
  [\href{https://arxiv.org/abs/1310.8042}{{\ttfamily 1310.8042}}].

\bibitem{Fuchs:2020cmm}
E.~Fuchs, O.~Matsedonskyi, I.~Savoray and M.~Schlaffer, \emph{{Collider
  searches for scalar singlets across lifetimes}},
  \href{https://doi.org/10.1007/JHEP04(2021)019}{\emph{JHEP} {\bfseries 04}
  (2021) 019}, [\href{https://arxiv.org/abs/2008.12773}{{\ttfamily
  2008.12773}}].

\bibitem{Carena:2022yvx}
M.~Carena, J.~Kozaczuk, Z.~Liu, T.~Ou, M.~J. Ramsey-Musolf, J.~Shelton et~al.,
  \emph{{Probing the Electroweak Phase Transition with Exotic Higgs Decays}},
  \href{https://doi.org/10.31526/lhep.2023.432}{\emph{LHEP} {\bfseries 2023}
  (2023) 432}, [\href{https://arxiv.org/abs/2203.08206}{{\ttfamily
  2203.08206}}].

\bibitem{ATLAS:2019nkf}
{\scshape ATLAS} collaboration, G.~Aad et~al., \emph{{Combined measurements of
  Higgs boson production and decay using up to $80$ fb$^{-1}$ of proton-proton
  collision data at $\sqrt{s}=$ 13 TeV collected with the ATLAS experiment}},
  \href{https://doi.org/10.1103/PhysRevD.101.012002}{\emph{Phys. Rev. D}
  {\bfseries 101} (2020) 012002},
  [\href{https://arxiv.org/abs/1909.02845}{{\ttfamily 1909.02845}}].

\bibitem{Fradette:2018hhl}
A.~Fradette, M.~Pospelov, J.~Pradler and A.~Ritz, \emph{{Cosmological beam
  dump: constraints on dark scalars mixed with the Higgs boson}},
  \href{https://doi.org/10.1103/PhysRevD.99.075004}{\emph{Phys. Rev. D}
  {\bfseries 99} (2019) 075004},
  [\href{https://arxiv.org/abs/1812.07585}{{\ttfamily 1812.07585}}].

\bibitem{Turner:1987by}
M.~S. Turner, \emph{{Axions from SN 1987a}},
  \href{https://doi.org/10.1103/PhysRevLett.60.1797}{\emph{Phys. Rev. Lett.}
  {\bfseries 60} (1988) 1797}.

\bibitem{Burrows:1988ah}
A.~Burrows, M.~S. Turner and R.~P. Brinkmann, \emph{{Axions and SN 1987a}},
  \href{https://doi.org/10.1103/PhysRevD.39.1020}{\emph{Phys. Rev. D}
  {\bfseries 39} (1989) 1020}.

\bibitem{Hardy:2024gwy}
E.~Hardy, A.~Sokolov and H.~Stubbs, \emph{{Supernova bounds on new scalars from
  resonant and soft emission}},
  \href{https://arxiv.org/abs/2410.17347}{{\ttfamily 2410.17347}}.

\bibitem{Caputo:2022mah}
A.~Caputo, H.-T. Janka, G.~Raffelt and E.~Vitagliano, \emph{{Low-Energy
  Supernovae Severely Constrain Radiative Particle Decays}},
  \href{https://doi.org/10.1103/PhysRevLett.128.221103}{\emph{Phys. Rev. Lett.}
  {\bfseries 128} (2022) 221103},
  [\href{https://arxiv.org/abs/2201.09890}{{\ttfamily 2201.09890}}].

\bibitem{Diamond:2023cto}
M.~Diamond, D.~F.~G. Fiorillo, G.~Marques-Tavares, I.~Tamborra and
  E.~Vitagliano, \emph{{Multimessenger Constraints on Radiatively Decaying
  Axions from GW170817}},
  \href{https://doi.org/10.1103/PhysRevLett.132.101004}{\emph{Phys. Rev. Lett.}
  {\bfseries 132} (2024) 101004},
  [\href{https://arxiv.org/abs/2305.10327}{{\ttfamily 2305.10327}}].

\end{thebibliography}\endgroup

\end{document}